\documentclass[nofloatfix,pre,onecolumn,reprint,aps,superscriptaddress,longbibliography]{revtex4-1}
\usepackage{feynmf}
\usepackage{graphicx}%
\usepackage{dcolumn}%
\usepackage{bm}%
\usepackage{color}
\usepackage[normalem]{ulem}
\usepackage{amsmath}
\usepackage{enumerate}
\usepackage{amsfonts}
\usepackage{epsfig}
\usepackage{graphicx}
\usepackage{cancel}
\usepackage[caption=false]{subfig}
\usepackage{floatrow}
\newcommand{\be}{\begin{equation}}
\newcommand{\ee}{\end{equation}}
\newcommand{\bea}{\begin{eqnarray}}
\newcommand{\eea}{\end{eqnarray}}
\usepackage{url}

\definecolor{green}{rgb}{0.0, 0.44, 0.0}
\definecolor{red}{rgb}{1.0, 0.13, 0.32}
\definecolor{blue}{rgb}{0.06, 0.2, 0.65}
\definecolor{darkgreen}{rgb}{0,0.5,0}
\definecolor{darkblue}{rgb}{0,0,0.6}
\definecolor{purple}{rgb}{0.4,.2,0.7}
\definecolor{magenta}{rgb}{1.0,0.0,1.0}
\usepackage[colorlinks=true,citecolor=darkgreen,linkcolor=purple,urlcolor=purple]{hyperref}

\def\le{\left}
\def\ri{\right}
\def\Ns{N}
\def\NJ{N_{\rm D}}
\def\Nn{N_{\rm n}}
\def\n{n}
\def\psuedo{{\cal G}}

\begin{document}

\title{Cycle-expansion method for the Lyapunov exponent, susceptibility, and higher moments}

\author{Patrick Charbonneau} 
\affiliation{Department of Chemistry, Duke University, Durham, North Carolina 27708, USA}
\affiliation{Department of Physics, Duke University, Durham, North Carolina 27708, USA}

\author{Yue (Cathy) Li} 
\affiliation{Department of Chemistry, Duke University, Durham, North Carolina 27708, USA}
\affiliation{Department of Electrical and Computer Engineering, Duke University, Durham, North Carolina 27708, USA}

\author{Henry D. Pfister}
\affiliation{Department of Electrical and Computer Engineering, Duke University, Durham, North Carolina 27708, USA}

\author{Sho Yaida}
\email{sho.yaida@duke.edu}
\affiliation{Department of Chemistry, Duke University, Durham, North Carolina 27708, USA}

\date{\today}

\begin{abstract}
Lyapunov exponents characterize the chaotic nature of dynamical systems by quantifying the growth rate of uncertainty associated with the imperfect measurement of initial conditions. Finite-time estimates of the exponent, however, experience fluctuations due to both the initial condition and the stochastic nature of the dynamical path. The scale of these fluctuations is governed by the Lyapunov susceptibility, the finiteness of which typically provides a sufficient condition for the law of large numbers to apply. Here, we obtain a formally exact expression for this susceptibility in terms of the Ruelle dynamical zeta function for one-dimensional systems. We further show that, for systems governed by sequences of random matrices, the cycle expansion of the zeta function enables systematic computations of the Lyapunov susceptibility and of its higher-moment generalizations. The method is here applied to a class of dynamical models that maps to static disordered spin chains with interactions stretching over a varying distance, and is tested against Monte Carlo simulations.
\end{abstract}

\maketitle

\section{\label{sec:level1}Introduction}
The Lyapunov exponent was initially devised to quantify the rate at which information dissipates in a chaotic dynamical system~\cite{Lyapunov,PP16}.
More concretely, it measures how the distance between two nearby trajectories scales exponentially with time when their initial conditions are infinitesimally close. The quantity has since found a number of other applications. For instance, it gives the free-energy density of one-dimensional spin chains~\cite{KW41} and the entropy rate of stationary hidden Markov models in information theory~\cite{Blackwell57,Pfister-2003,Holliday-it06,JSS08}. Interest in Lyapunov exponents continues to spread, as illustrated by the recent study of black-hole scrambling~\cite{HP07,SS08}, which results in the formulation of an upper bound on the Lyapunov exponent for quantum systems~\cite{MSS16} (see also Ref.~\cite{Kurchan16}).

When the dynamics of a system can be modeled by a sequence of randomly-drawn matrices~\cite{CPV12}, the Lyapunov exponent is also intimately connected to the rich properties of disordered systems~\cite{MPV87}. For the sake of concreteness, consider a sequence of matrices, $\{T_i\}_{i\in \mathbb{N}}$, be they transfer matrices in disordered spin chains or transition-observation matrices in hidden Markov chains~\cite{Pfister-2003}. The Lyapunov exponent is then the typical growth rate of the maximum-modulus eigenvalue of the product of these matrices. More formally, we define the finite-sample quantity
\be \label{lyap}
\lambda_{N}^{(\alpha)}\equiv\frac{1}{\Ns}\ln\le(\le\| \prod_{i=1}^{\Ns} T_i^{(\alpha)}\ri\|\ri)\, ,
\ee
where $\Ns$ is the sample size and $\alpha$ denotes the sequence of random matrices drawn independently from some fixed underlying probability distribution.
The Lyapunov exponent is then given by the infinite system size limit
\begin{equation}\label{defL}
\lambda\equiv\lim_{\Ns\rightarrow\infty}\mathbb{E}[\lambda_{\Ns}]\equiv\lim_{\Ns\rightarrow\infty}\lim_{\NJ\rightarrow\infty}\frac{1}{\NJ}\sum_{\alpha=1}^{\NJ}\lambda_{N}^{(\alpha)}\, ,
\end{equation}
where the disorder-average $\mathbb{E}[...]$ can be obtained by drawing $\NJ$ disorder realizations.
Note that the limit does not depend on the choice of matrix norm, $\|\ldots\|$.

Given the ubiquitous appearance of the Lyapunov exponent in products of random matrices~\cite{Nielsen97,CPV12}, many methods have been developed for its estimation, including Monte Carlo algorithms~\cite{BGS76,BGGS80,Vanneste10}, a perturbative weak-disorder expansion~\cite{GID84,DG84}, a microcanonical method~\cite{DP89}, a cycle expansion~\cite{Mainieri92}, a Dyson-Schmidt equation~\cite{DH83,WM96}, a scaling method~\cite{PS92,Davids94}, an evolution-operator method~\cite{Bai07}, and an infinite transfer matrix method~\cite{Bai09}.
Central to all these approaches is the assumption that sample-to-sample fluctuations of $\lambda_N$ are not so large as to invalidate the law of large numbers. Interestingly, in assessing the applicability of this law, an essential role is played by the second moment of the generalized Lyapunov exponent~\cite{Fujisaka83}, i.e., the Lyapunov susceptibility,
\be\label{LS}
\chi_{\rm L}\equiv \lim_{\Ns\rightarrow\infty} \Ns\le(\mathbb{E}[\lambda_{\Ns}^2]-\mathbb{E}[\lambda_{\Ns}]^2\ri)\, .\nonumber
\ee
It has indeed been proven under certain conditions on the underlying matrix distribution that the central limit theorem holds if and only if $\chi_{\rm L}$ is finite~\cite{Ishitani77}, thus providing a sufficient (though not necessary) condition for the law of large numbers to hold. 
The susceptibility also appears in rigorous treatments of mean-field spin-glasses~\cite{ALR87,BL16} and is related to bond chaos~\cite{BM87,BKM03,Aspelmeier08}. For turbulent flows, a nontrivial susceptibility further signals the existence of intermittency~\cite{BPPV85,Frisch95}.
Given the physical and mathematical importance of this quantity~\cite{CCPVV14}, it is surprising that it has thus far rarely been explicitly considered. 

Here, we develop methods for evaluating the Lyapunov susceptibility, and use the results to understand better its behavior. More specifically, we extend the cycle-expansion method~\cite{Cvitanovic88,AAC90,Mainieri92}, which is based on the Ruelle dynamical zeta function and provides a formally exact expression linking the underlying cycles to the susceptibility. We further find that, when applicable, the cycle-expansion method offers a natural and efficient approach for assessing tails of the Lyapunov-exponent distribution pertaining to the physics of large deviations.

The rest of this paper is organized as follows. In Sec.~\ref{sec:models} concrete models that we use to illustrate our methodology are stipulated. Results of Monte Carlo simulations are discussed in Sec.~\ref{sec:MC}, and the cycle-expansion method for the Lyapunov susceptibility and its higher-moment generalizations is developed in Sec.~\ref{sec:ReplicaCycle}. In Sec.~\ref{sec:compare} results of cycle expansions are compared against those of Monte Carlo simulations. A brief conclusion follows in Sec.~\ref{sec:conclusion}.

\section{\label{sec:models}Models}
This section introduces the class of models used in the rest of this work. The models consist of a static one-dimensional chain of $\Ns$ spins with an interaction of range $\Nn$ captured by transfer matrices, $T_a$. They thus constitute a generic set of one-dimensional disordered models with finite-range interactions. They can equivalently be viewed as $\Nn$-neighboring spins that evolve dynamically with a transition matrix $T_a$ hitting at each time step, or as a single spin evolving with finite-time memory. It is worth stressing, however, that these models are chosen mainly for illustrative purposes, and that the methods developed below have a much broader scope of application. 

\subsection{Nearest-neighbor (NN) model}
The disordered NN Ising model is governed by the Hamiltonian
\be
H=-\sum_{i=1}^{\Ns}J_i S_iS_{i+1},
\ee
where spins $S_i=\pm1$ for $i=1,\ldots,\Ns$ with periodic boundary condition $S_{\Ns+1}=S_1$.
The NN interactions $\{J_i\}_{i=1,\ldots,\Ns}$ are randomly drawn to be $\pm J$ with equal probability $1/2$ at each site. The associated transfer matrices, $T^{(\alpha)}_i\in\le\{T_{+}, T_{-}\ri\}$, are then
\be\label{TNN}
T_{\pm}\equiv\begin{bmatrix}
e^{\pm\beta J} & e^{\mp\beta J} \\ e^{\mp\beta J} & e^{\pm\beta J}\, 
\end{bmatrix}
=\begin{bmatrix}
e^{\pm\tilde{\beta}} & e^{\mp\tilde{\beta}} \\ e^{\mp\tilde{\beta}} & e^{\pm\tilde{\beta}}\, 
\end{bmatrix}
\ee
for the dimensionless inverse temperature $\tilde{\beta}\equiv\beta J\equiv \frac{J}{k_{\rm B}T}$.
The free-energy density,
\be\label{fed}
\beta f_N^{(\alpha)}=-\frac{1}{N}\ln\le[\text{tr}\le(\prod_{i=1}^N T_i^{(\alpha)}\ri)\ri]\, ,
\ee
is thus related to the associated Lyapunov exponent in the thermodynamic limit $\Ns\rightarrow\infty$ through Gelfand's formula~\cite{Gelfand41}, i.e., $\lim_{\Ns\rightarrow\infty}\lambda_{\Ns}^{(\alpha)}=\lim_{\Ns\rightarrow\infty}\le(-\beta f_{\Ns}^{(\alpha)}\ri)$. For this particular model, each disorder realization can be mapped onto a pure Ising model without disorder by redefining the spins (combined with the possible replacement of periodicity by antiperiodicity at the boundary),
and hence is fully solvable, with
\be
\lambda=-\beta f=\ln[2\cosh(\tilde{\beta})]\, 
\ee
and $\chi_{\rm L}=0$.

\subsection{Next-nearest-neighbor (NNN) model}
Including NNN interactions is sufficient to make the analysis nontrivial. The Hamiltonian is then
\be
H=-\sum_i \le(J_i^{[1]} S_i S_{i+1} + J_i^{[2]} S_i S_{i+2}\ri)\, ,
\ee
where $J_i^{[l]}=\pm\frac{J}{\sqrt{2}}$ are independent and identically distributed random variables with amplitude chosen such that the NN scaling of the Lyapunov exponent,
\be\label{Jnormalization}
\lambda=\ln(2)+\frac{\tilde{\beta}^2}{2}+O(\tilde{\beta}^4)\, ,
\ee
is recovered at high temperatures.
This model has four possible transfer matrices~\cite{SF79,footnote_Ishitani}
\begin{widetext}
\begin{align}\label{NNNtransmat}
T_{\le(J^{[1]},J^{[2]}\ri)}\equiv\begin{bmatrix}
e^{\beta(J^{[1]}+J^{[2]})} & e^{\beta(J^{[1]}-J^{[2]})} & 0 & 0\\
0 & 0 & e^{\beta(-J^{[1]}+J^{[2]})} & e^{\beta(-J^{[1]}-J^{[2]})}\\
e^{\beta(-J^{[1]}-J^{[2]})} & e^{\beta(-J^{[1]}+J^{[2]})} & 0 & 0\\
0 & 0 & e^{\beta(J^{[1]}-J^{[2]})} & e^{\beta(J^{[1]}+J^{[2]})}
\end{bmatrix}
\end{align}
\end{widetext}
that occur with equal probability, $1/4$. The Lyapunov exponent is here again related to the free-energy density through Gelfand's formula. The model, however, cannot be mapped to a solvable nondisordered model because of the frustration generically introduced by conflicting NN and NNN couplings.

\subsection{Generalized nearest-neighbor models}
The generalization of these models to $\Nn$ nearest neighbors,
\be
H=-\sum_i \le(\sum_{l=1}^{\Nn}J_i^{[l]} S_i S_{i+l} \ri)\, 
\ee
with $J_i^{[l]}=\pm \frac{J}{\sqrt{\Nn}}$, results in $2^{\Nn}$ equally probable $2^{\Nn}$-by-$2^{\Nn}$ transfer matrices with elements
\bea\label{Tgen}
&&\le[T_{\le(J^{[1]},\ldots,J^{[\Nn]}\ri)}\ri]_{(\sigma_1,\ldots,\sigma_{\Nn}),(\sigma'_2, \ldots,\sigma'_{\Nn+1})}\\
&\equiv&\delta_{\sigma_2,\sigma'_2}\cdots\delta_{\sigma_{\Nn},\sigma'_{\Nn}}\exp\le[\beta\sum_{l=1}^{\Nn}J^{[l]} \sigma_1  \sigma'_{1+l}\ri]\nonumber ,
\eea
where the dummy spin variables $\sigma_k=\pm1$ and $\sigma'_k=\pm1$ span the $2^{\Nn}$-dimensional vector space. Note that $\Nn=1$ recovers the NN model and $\Nn=2$ the NNN model, while the limit $\Nn\rightarrow\infty$ corresponds to the Sherrington-Kirkpatrick model with $\pm J/\sqrt{\Ns}$ disorder (which, unlike the model with the canonical Gaussian form~\cite{SK75}, has not been solved in the literature). This model therefore offers yet another way of interpolating between finite- and infinite-dimensional systems~\cite{KAS83,AWMK16,FJP09,CP10,CY17}.

\section{\label{sec:MC}Monte Carlo simulations}
The Lyapunov exponent [Eq.~\eqref{lyap}] and its susceptibility [Eq.~\eqref{LS}] for the above models can be directly evaluated by computing the largest eigenvalue for the product of each sequence of matrices.
Because such computation for a large number of matrices results in numerical inaccuracy, any reasonable implementation cannot apply this scheme directly, but instead keeps track of the growth rate of the vector magnitude~\cite{BGS76,BGGS80,Skokos10}.
More specifically, we here randomly pick an initial normalized $2^{\Nn}$-dimensional vector, $\mathbf{v}(i=1)$, evaluate the magnification factor $m^{(\alpha)}_i=\| T^{(\alpha)}_i \mathbf{v}(i)\|$ after each transfer matrix multiplication, and then define a new normalized vector, $\mathbf{v}(i+1)\equiv T^{(\alpha)}_i \mathbf{v}(i)/m^{(\alpha)}_i$. In order to lose memory of the arbitrarily chosen initial vector, the first $N_{\rm equi}$ equilibration steps are discarded, hence the estimate for the sample Lyapunov exponent is
\be
\lambda^{(\alpha)}_N=\frac{1}{\Ns}\sum_{i=N_{\rm equi}+1}^{\Ns+N_{\rm equi}}\ln\le(m^{(\alpha)}_i\ri)\, .
\ee
Averaging over $\NJ$ samples provides an estimate of the Lyapunov exponent, while computing the sample variance yields the Lyapunov susceptibility upon proper normalization with $\Ns$.
This scheme can be further generalized to extract the second-largest eigenvalue of the product of random matrices, $\lambda_{\rm sub}$, through Gram-Schmidt orthogonalization~\cite{BGGS80}. This second eigenvalue encodes the correlation length (or correlation time from a dynamical viewpoint), $\xi\equiv1/(\lambda-\lambda_{\rm sub})$.
We here obtain results with $N_{\rm equi}=10^5$, $\Ns=10^6$,  and $\NJ=10^5$. In particular, the equilibration time $N_{\rm equi}$ is chosen to be much longer than the correlation length/time at all temperatures considered.

\begin{figure*}[t]
\centerline{
\hspace{-1.0in}
\subfloat[]{\includegraphics[width=0.315\textwidth]{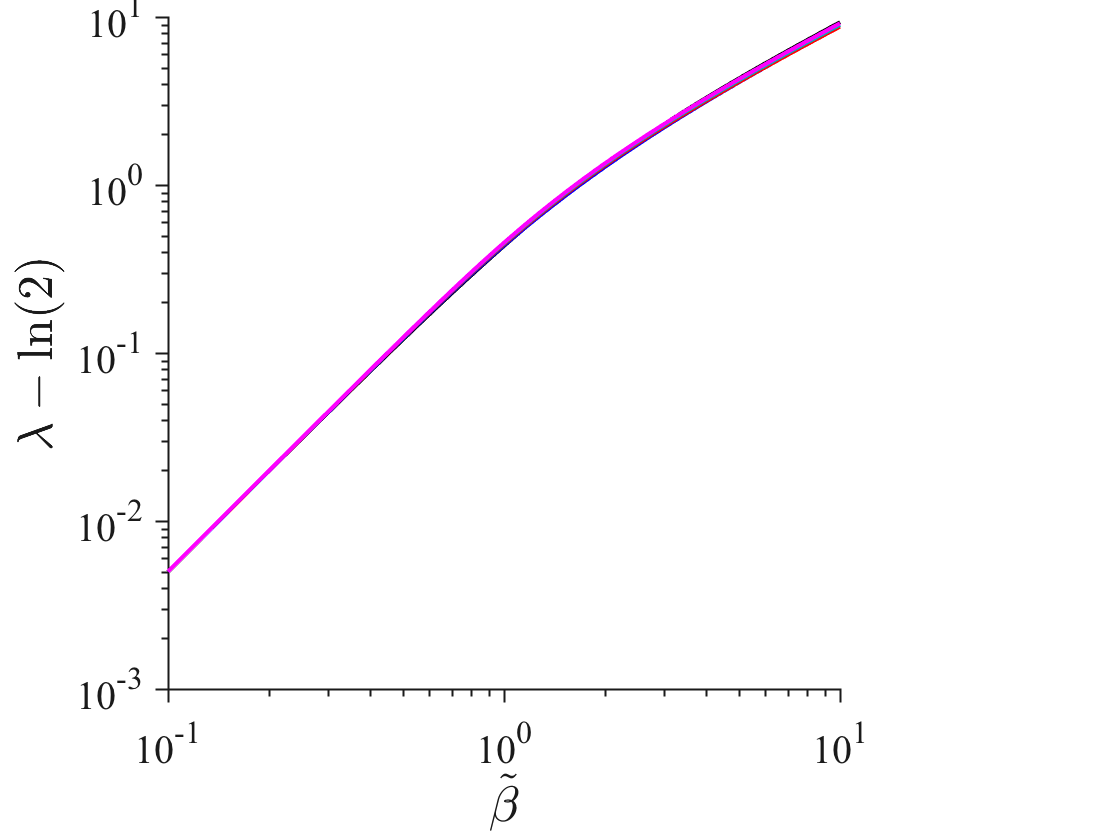}}\quad%
\hspace{-0.7in}
\subfloat[]{\includegraphics[width=0.315\textwidth]{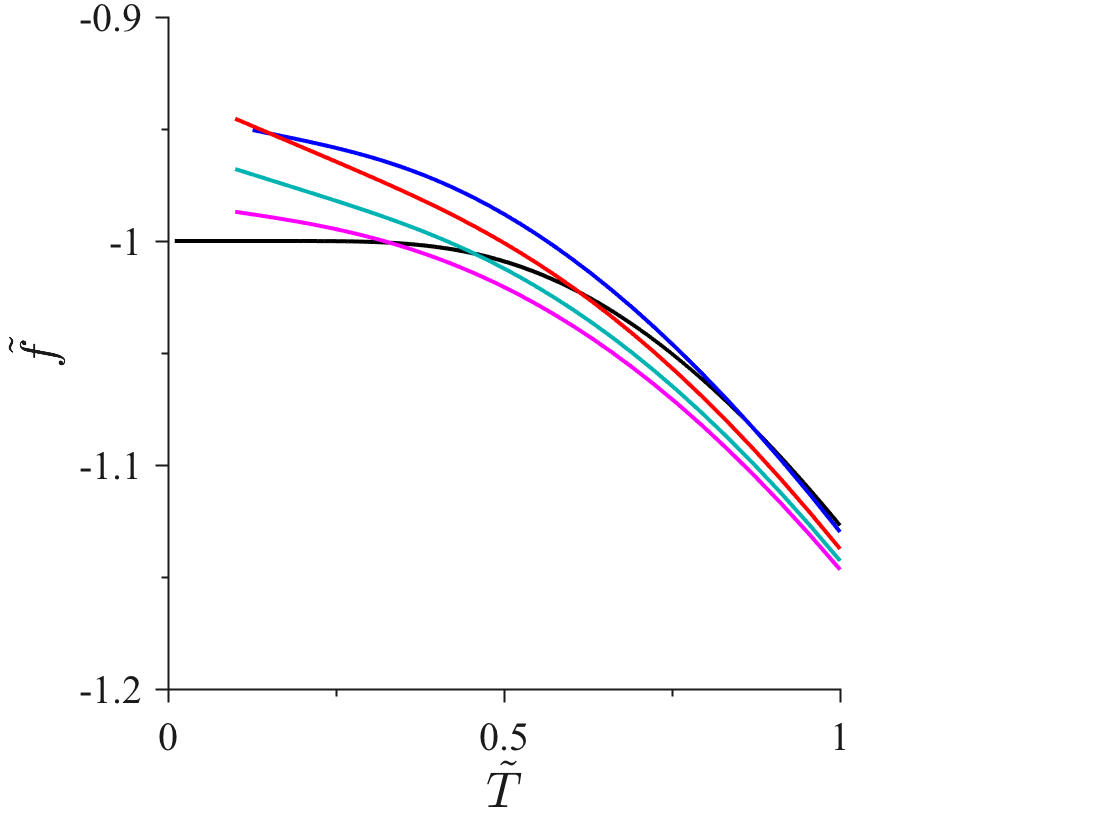}}\quad%
\hspace{-0.7in}
\subfloat[]{\includegraphics[width=0.315\textwidth]{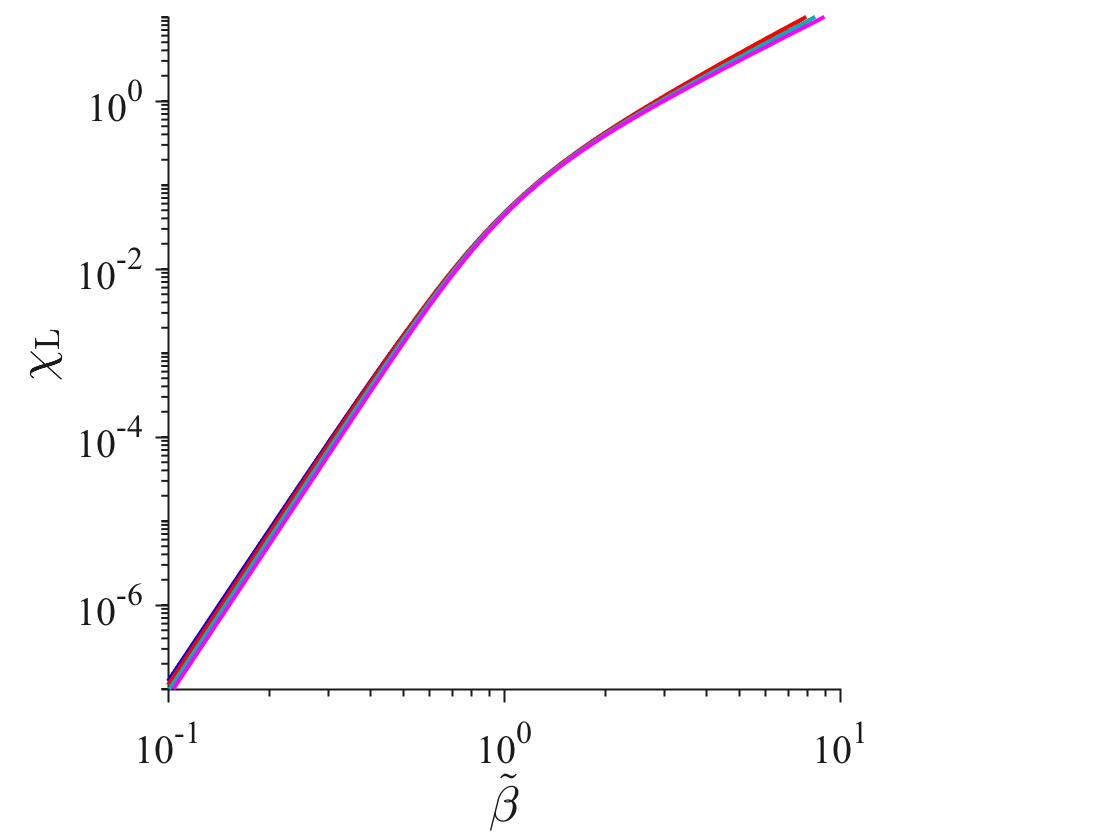}}\quad%
\hspace{-0.7in}
\subfloat[]{\includegraphics[width=0.315\textwidth]{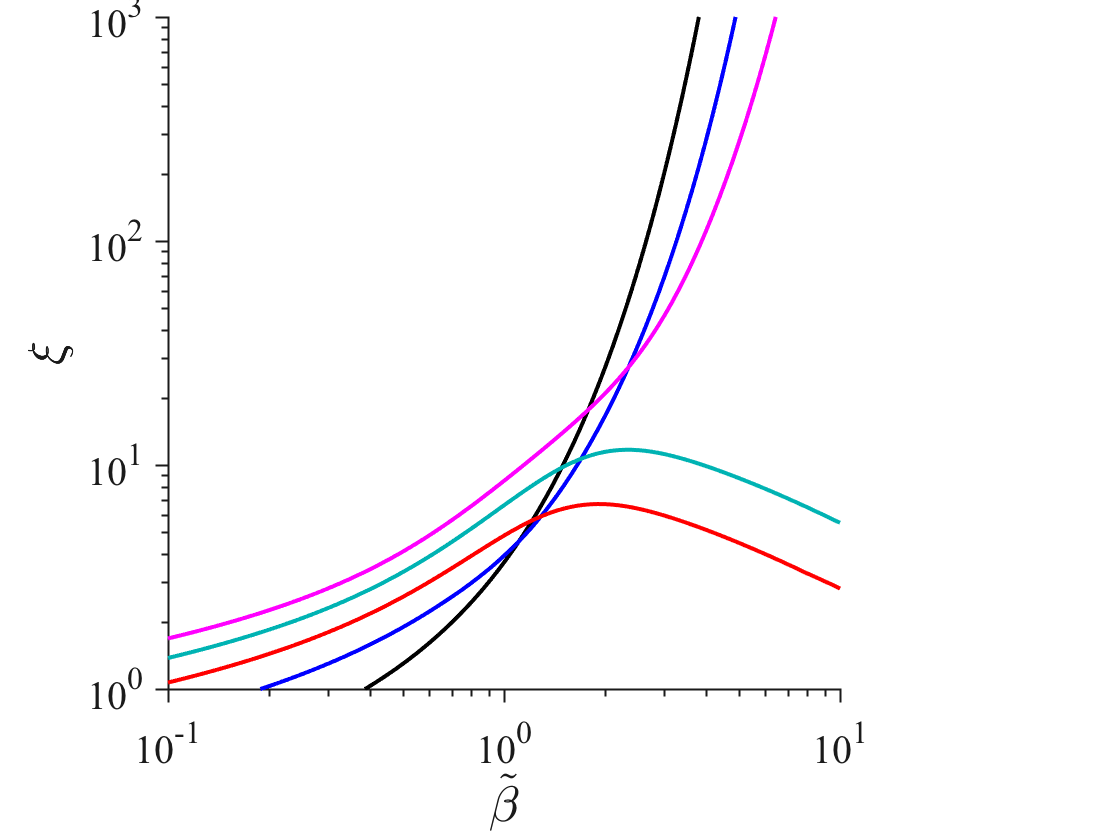}}
\hspace{-1in}
}
\caption{Temperature dependence of the Lyapunov observables obtained by Monte Carlo simulations for $\Nn=2$ (blue), $3$ (red), $4$ (cyan), and $5$ (magenta), along with the exact results for $\Nn=1$ (black). (a) Subtracting the high-temperature limit, $\ln(2)$, from $\lambda$ shows it to scale as $\tilde{\beta}^2$ for $\tilde{\beta}\ll1$ and as $\tilde{\beta}$ for $\tilde{\beta}\gg1$. (b) The normalized free-energy density $\tilde{f}\equiv f/J$ as a function of the normalized temperature $\tilde{T}\equiv T/J$. For $\Nn\geq3$, the normalized free-energy monotonically varies as the interaction range increases. It is expected to asymptote to the mean-field Sherrington-Kirkpatrick limit with $\pm J/\sqrt{\Ns}$ disorder distribution, but such a disordered model has not yet been solved. (c) The Lyapunov susceptibility, $\chi_{\rm L}$, also shows two regimes. Its growth as $\tilde{\beta}^2$ for $\tilde{\beta}\gg1$ reveals the finiteness of $\lim_{\Ns\rightarrow\infty} \Ns\le(\mathbb{E}[f_{\Ns}^2]-\mathbb{E}[f_{\Ns}]^2\ri)$, which governs sample-to-sample fluctuations in the thermal  free energy. (d) The correlation length, $\xi$, diverges exponentially toward $T=0$ for $\Nn=1$, $2$, and $5$, but is maximal at $\tilde{T}\sim0.5$ for $\Nn=3$ and $4$, thus demonstrating the nonuniversality of the appropriate order parameter across models of disordered spin chains. Note that for all the temperatures considered, $\xi<N_{\mathrm{equil}}=0.1\Ns$. Note also that the error on the various quantities is of order $\sqrt{\chi_{\rm L}/(\NJ\Ns)}$, and is thus much smaller than the width of the lines in this figure.
}
\label{LyapunovO}
\end{figure*}

Results for $\lambda$ and $\chi_{\rm L}$ are given in Figs.~\ref{LyapunovO}(a) and~\ref{LyapunovO}(c), respectively, for $\Nn=1,\ldots,5$. They all show the same qualitative trend. At high temperatures, the Lyapunov exponent is well described by Eq.~\eqref{Jnormalization} with $\lambda(T\rightarrow\infty)=\ln(2)$, the entropy density of noninteracting spins; at low temperatures, $\lambda\sim\tilde{\beta}$, which is consistent with the free energy approaching a constant at $T=0$ [Fig.~\ref{LyapunovO}(b)]. In that same limit, the susceptibility scales as $\chi_{\rm L}\sim\tilde{\beta}^2$ , which suggests that sample-to-sample fluctuations in the free-energy density are $O(\Ns^{-1/2})$. Hence with our choice of $\Ns$ and $\NJ$, the estimates of the Lyapunov exponent have an accuracy roughly of one part in one hundred thousand, which is much smaller than the thickness of the lines in Fig.~\ref{LyapunovO}.

The correlation length, $\xi$, is reported in Fig.~\ref{LyapunovO}(d). For $\Nn=2$ and $\Nn=5$, the length grows exponentially toward $T=0$, just as in the pure Ising model with $\Nn=1$. For $\Nn=3$ and $4$, by contrast, $\xi$ initially grows upon cooling but then decays, reaching a maximum around $T/J\sim 0.5$.
This result may seem surprising at first, but in fact reflects the subtlety of defining order 
parameters. The relevant quantity depends on the details of the microscopic interactions and is thus nonuniversal across models~\cite{Baxter71,BW74,KM77,Barber79}. As an illustration, consider the Sherrington-Kirkpatrick $\Nn\rightarrow\infty$ limit, for which ordering is of a completely different (amorphous) nature. In order to capture amorphous ordering at low temperatures, correlation functions have to be appropriately modified. It should therefore not be surprising that $\xi=1/(\lambda-\lambda_{\rm sub})$ associated with a particular $\Nn$-spin correlation function does not exhibit a low-temperature divergence for some of the models intermediate between $\Nn=1$ and $\Nn=\infty$. 

\section{\label{sec:ReplicaCycle}Replica trick}
In this section, we first review the use of the replica trick to average over disorder, and then develop the cycle-expansion method. We show below how the latter is closely related to the former, but surmounts some of its implementation difficulties.

\subsection{Replica trick}
For systems with quenched disorder, the Lyapunov exponent in Eq.~\eqref{lyap} involves averaging over a logarithm, which is often analytically intractable. The replica trick sidesteps this problem by looking instead at~\cite{HLP34,EA75,MPV87}
\bea
L(\n)&\equiv&\lim_{N\rightarrow\infty}L_N(\n)\equiv\lim_{N\rightarrow\infty}\frac{1}{N}\ln\left\{\mathbb{E}\le[ e^{\n N\lambda_N}\ri] \right\}\nonumber\\
&=&\lim_{N\rightarrow\infty}\frac{1}{N}\ln\left\{\mathbb{E}\left[ \le\|\prod_{i=1}^N T_i\ri\|^{\n} \right] \right\}\, ,
\eea
which, for integer $\n$, can be regarded as the logarithm of the average over $\n$ replicated samples~\cite{Pendry82,BGHLM86,OP96}. This quantity is both analytically and computationally more tractable. The Lyapunov exponent can then be obtained as
\bea\label{deriva}
\frac{\mathrm{d}{L(\n)}}{\mathrm{d} \n}\Bigg|_{\n=0}&=&\lim_{\n\rightarrow0}\lim_{N\rightarrow\infty}\frac{\mathrm{d}{L_N(\n)}}{\mathrm{d} \n}   \nonumber\\
&=&\lim_{\n\rightarrow0}\lim_{N\rightarrow\infty}\frac{\mathbb{E}\left[e^{\n N\lambda_N}\lambda_N\right]}{\mathbb{E}\left[e^{\n N\lambda_N}\right]} \nonumber\\
&=&\lim_{N\rightarrow\infty}\mathbb{E}\left[\lambda_N\right]=\lambda \, ,
\eea
assuming that the order of the limits over $N$ and $\n$ can be swapped, which is not (yet) a mathematically rigorous step~\cite{HP79}. It is possible, however, to rigorously establish that both limits exist and that $L'(0)\geq \lambda$ (see Appendix~\ref{sec:prop_gen_lya}).
A similar computation and set of assumptions yield the Lyapunov susceptibility~\cite{footnote_large}
\be
\chi_{\rm L}=\frac{\mathrm{d}^2{L(\n)}}{\mathrm{d} \n^2}\Bigg|_{\n=0}\, .
\ee
We next assume that the generalized Lyapunov exponent $L(\n)$ is an analytic function for $\n\in[0,\infty)$. Although once again not rigorous, this hypothesis is physically reasonable. No transition--including a replica-symmetry-breaking transition--can indeed occur at finite temperature in one-dimensional systems with short-range interactions~\cite{Ruelle79}.

\begin{figure}
\centerline{
\includegraphics[width=0.8\textwidth]{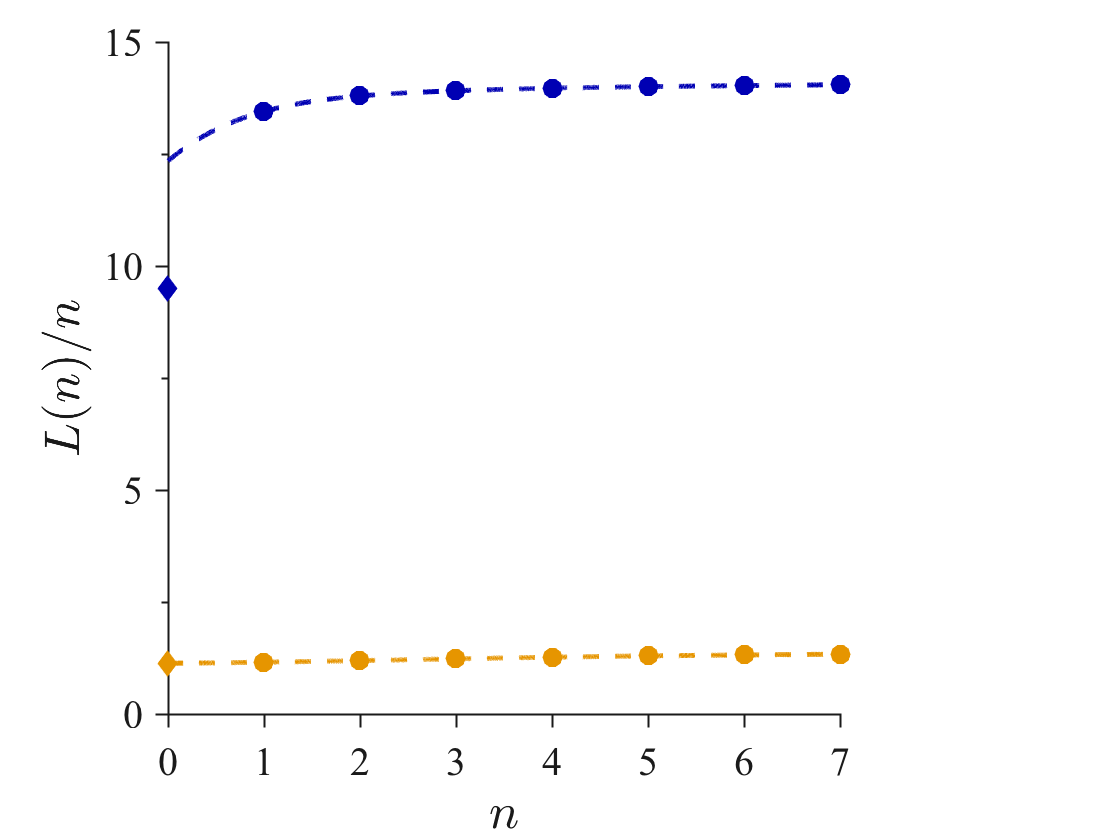}
}
\caption{Generalized Lyapunov exponents evaluated at integer values of $n$ for the NNN model at $\tilde{\beta}=1$ (orange) and $\tilde{\beta}=10$ (navy-blue). The diamonds at $\n=0$ denote the values for $\lim_{\n\rightarrow0}\frac{L(\n)}{\n}=\lambda$ obtained by Monte Carlo simulations. The values of the generalized Lyapunov exponent are obtained through the tensorial replication technique described in Refs.~\cite{BGHLM86,CPV12} for $\n=1,\ldots,7$. A sextic polynomial fit (dashed lines) gets more or less the correct result for $\tilde{\beta}=1$, but fails dramatically for $\tilde{\beta}=10$.}
\label{Lqq}
\end{figure}

Based on these results and assumptions, one might expect the Lyapunov exponent and susceptibility to be obtained by extrapolating the slope and curvature, respectively, of $L(\n)$ computed at positive integer $\n$ to $\n=0$~\cite{BGHLM86}. Specifically, the assumed analyticity permits a Taylor expansion
\begin{align}\label{Taylor}
L(\n) &=L(0)+L'(0)\n+\frac{1}{2}L''(0)\n^2+\frac{1}{6}L'''(0)\n^3+\ldots\, 
\end{align}
with $L(0)=0$ (by definition), $L'(0)=\lambda$, and $L''(0)=\chi_{\rm L}$.  Figure~\ref{Lqq}, however, makes clear the technical difficulty of such an extrapolation. As discussed in Sec.~\ref{sec:MC}, models with $\Nn\geq2$ in the low temperature regime, $\tilde{\beta}\gg 1$, have $\chi_L\propto\tilde{\beta}^2$, while $\lambda\propto\tilde{\beta}$. As a result, $\frac{L(\n)}{\n}$ dips quickly as $n$ approaches the origin. This rapid curbing prevents the reliable extrapolation of the intercept from function evaluations at positive integer $\n$, even if these evaluations are obtained with machine precision and elaborate extrapolation schemes, such as Pad\'{e} approximants, are used. In practice, at low temperatures such a scheme simply fails. 

In passing, we note that the replica trick can also be used to recover an exact integral equation for the Lyapunov exponent~\cite{DH83,WM96}.
In order to attain the accuracy of order $\epsilon$ through such a scheme, however, the computational cost scales as $\le(1/\epsilon\ri)^{m-1}$ for $m$-by-$m$ transfer matrices due to the need for discretizing the interval of length $m$ into steps of size $\epsilon$. Thus for $m>3$ (i.e., for $\Nn>1$) this approach quickly becomes outperformed by the Monte Carlo algorithm, which has a computational cost that scales as $(1/\epsilon)^2$.

\subsection{Cycle expansion}
In order to avoid the numerical challenge of a direct replica extrapolation, we instead consider the cycle-expansion method~\cite{Cvitanovic88,AAC90,Mainieri92}. This computational scheme begins by constructing  the Ruelle dynamical zeta function~\cite{Ruelle78,Ruelle02},
\be
\zeta^{-1}\le(z,\n\ri)\equiv \exp\le\{-\sum_{\Ns=1}^{\infty} \frac{1}{\Ns}\le[z e^{L_{\Ns}(\n)}\ri]^{\Ns}\ri\}\, ,
\ee
which can be evaluated systematically by cycle expansion. Recall that $L_{\Ns}(0)=0$, hence for $\n=0$ the series that appears in the argument of the exponential can be explicitly summed to yield $\zeta^{-1}\le(z,0\ri)=1-z$, which has a zero at $z=1$. In general, given the thermodynamic limit $\lim_{\Ns\rightarrow\infty}L_{\Ns}(\n)=L(\n)$,
\be
\zeta^{-1}\le(e^{-L(\n)},\n\ri)=0\, .
\ee
Differentiating the above relation with respect to $\n$ then yields~\cite{footnote_Bai}
\bea
\lambda&=&L'(0)=-\partial_{\n}\zeta^{-1}\le(1,0\ri)\,\,\mathrm{and }\nonumber\\
\chi_{\rm L}&=&L''(0)=\lambda^2-\partial^2_{\n}\zeta^{-1}\le(1,0\ri)+2\lambda\partial_{z}\partial_{\n}\zeta^{-1}\le(1,0\ri)\, .\nonumber
\eea
Given the formal expression for the zeta function from the cycle expansion [as detailed in Appendix~\ref{CycleNotations}, $P^{\star}$ denotes the set of pseudocycles $\psuedo$, with the sign $\mathcal{M}(\psuedo)$, the probability $p(\psuedo)$, the length $\ell(\psuedo)$, and the spectral radius $\rho(\psuedo)$, i.e., (the product of) the largest eigenvalue(s)]
\be\label{RuelleGod}
\zeta^{-1}\le(z,\n\ri)=\sum_{\psuedo\in P^{\star}}\mathcal{M}(\psuedo)p(\psuedo)z^{\ell(\psuedo)} e^{\n \ln\rho(\psuedo)}\, ,
\ee
we first recover the expression for the Lyapunov exponent~\cite{Mainieri92,Nielsen97}
\be\label{cyclelyap}
\lambda=-\sum_{\psuedo\in P^{\star}}\mathcal{M}(\psuedo)p(\psuedo) \ln\rho(\psuedo)\, 
\ee
and then obtain the Lyapunov susceptibility
\be\label{cyclezeta}
\chi_{\rm L}=\lambda^2 +\sum_{\psuedo\in P^{\star}}\mathcal{M}(\psuedo)p(\psuedo) \ln\rho(\psuedo)\le[2\lambda \ell(\psuedo)-\ln\rho(\psuedo)\ri]\, .
\ee
Higher moments of the distribution for $\lambda_{\Ns}$ can also be obtained by further differentiating with respect to $\n$. For example, the third derivative is proportional to the skewness 
\bea
L'''(0)&=&\lim_{\Ns\rightarrow\infty} \Ns^2\le\{\mathbb{E}[\le(\lambda_{\Ns}-\mathbb{E}[\lambda_{\Ns}]\ri)^3]\ri\}\, \\
&=&-\lambda^3+3\chi_{\rm L}\lambda-\partial^3_{\n}\zeta^{-1}\le(1,0\ri)+3\lambda\partial_{z}\partial_{\n}^2\zeta^{-1}\le(1,0\ri)\, \nonumber\\
&&-3\lambda^2\partial_{z}^2\partial_{\n}\zeta^{-1}\le(1,0\ri)+3\le(\chi_{\rm L}-\lambda^2\ri)\partial_{z}\partial_{\n}\zeta^{-1}\le(1,0\ri)\, \nonumber
\eea
and the fourth derivative is proportional to the kurtosis 
\bea
&&L''''(0)\\
&=&\lim_{\Ns\rightarrow\infty} \Ns^3\le\{\mathbb{E}[\le(\lambda_{\Ns}-\mathbb{E}[\lambda_{\Ns}]\ri)^4]-3\mathbb{E}[\le(\lambda_{\Ns}-\mathbb{E}[\lambda_{\Ns}]\ri)^2]^2\ri\}\, .\nonumber
\eea
Because each differentiation brings down an overall factor of $\Ns$, higher-order derivatives are associated with ever refined information about the distribution. 

We can further generalize the cycle expansion to glean information about the whole Lyapunov characteristic exponent spectrum and, in particular, about the second largest eigenvalue that controls the correlation length. The derivation of the cycle-expansion expression for the zeta function indeed only depends on the positivity and cyclic nature of the weight. Hence the above formulae also provide the magnitude of the subleading eigenvalues via a straightforward replacement of the spectral radius, $\rho(\psuedo)$, by the magnitude of the corresponding rank eigenvalues.

\begin{figure*}[t]
\centerline{
\hspace{-1.0in}
\subfloat[]{\includegraphics[width=0.315\textwidth]{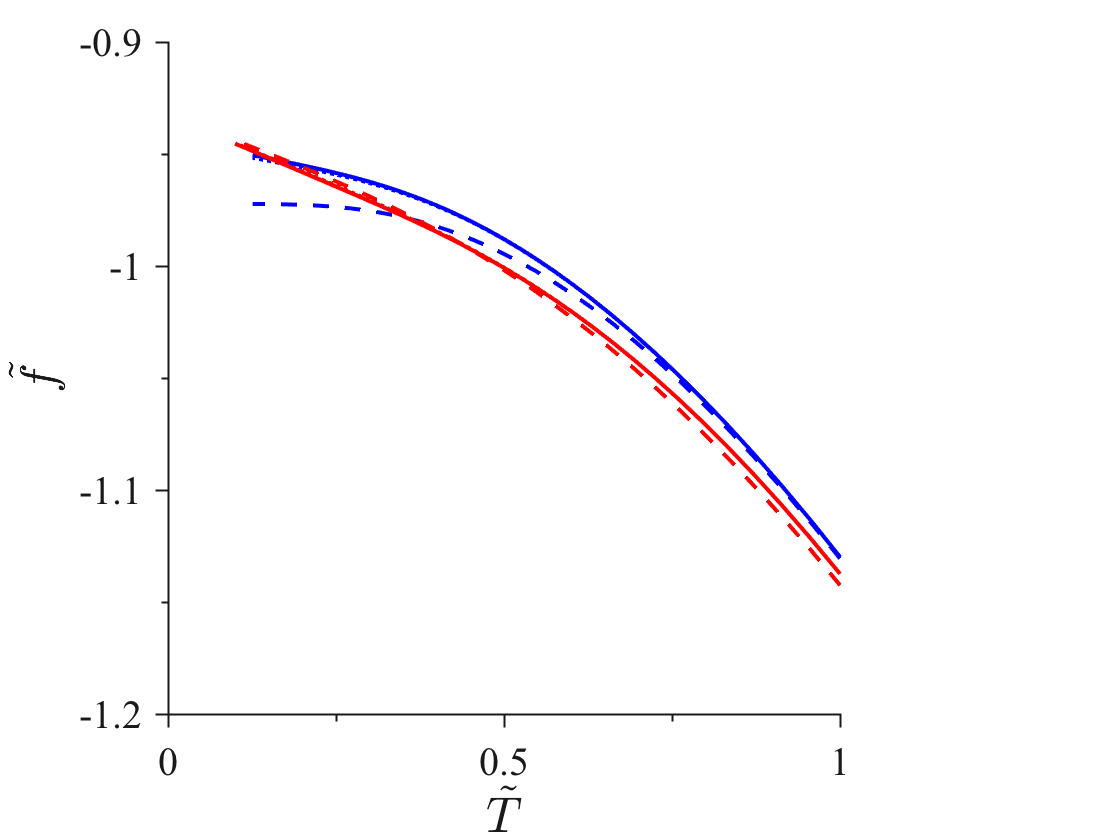}}\quad%
\hspace{-0.7in}
\subfloat[]{\includegraphics[width=0.315\textwidth]{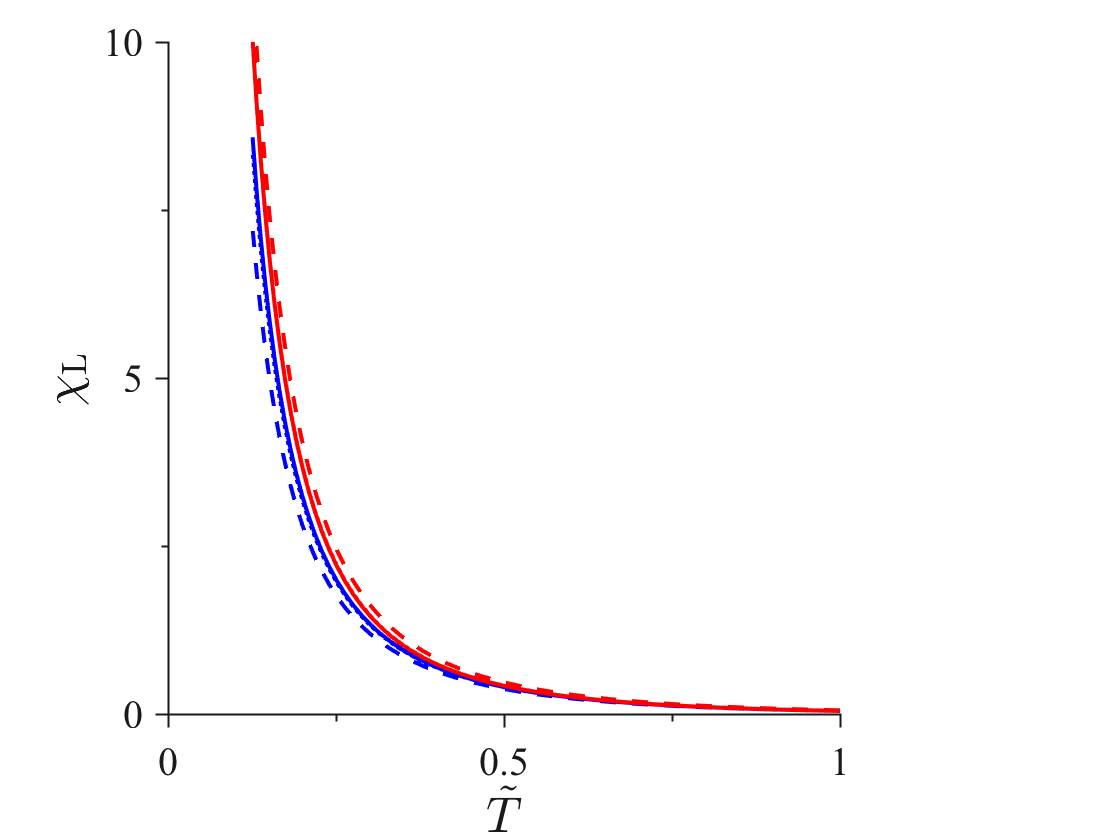}}\quad%
\hspace{-0.7in}
\subfloat[]{\includegraphics[width=0.315\textwidth]{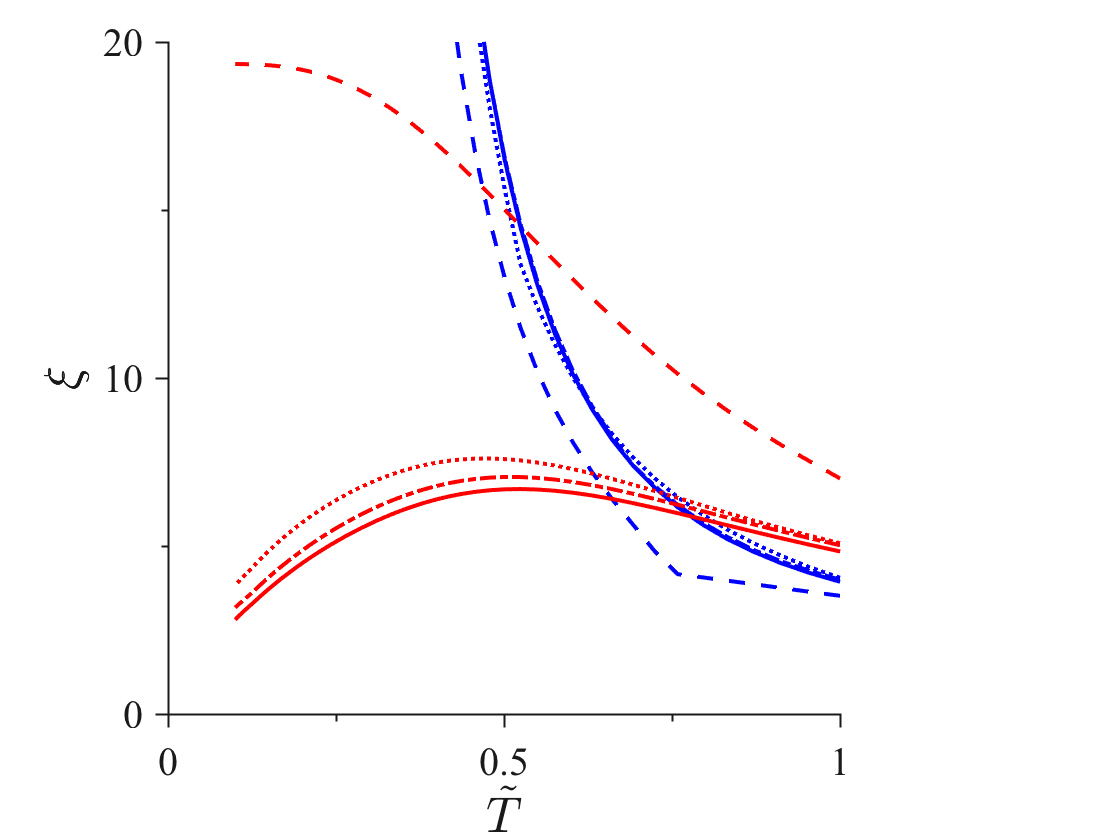}}
\hspace{-1in}
}
\caption{Lyapunov observables obtained through the cycle expansion method for $\Nn=2$ (blue) and $3$ (red) at order $k=3$ (dashed), 7 (dotted), and 11 (dash-dotted), along with the Monte Carlo simulations results (solid). (a) The normalized free-energy $\tilde{f}$ as a function of the normalized temperature $\tilde{T}$. (b) The Lyapunov susceptibility, $\chi_{\rm L}$, as a function of  $\tilde{T}$. (c) The correlation length, $\xi$, as a function of $\tilde{T}$. Convergence of the cycle expansion for the subleading eigenvalue is slower than for the leading eigenvalue, but nonetheless suffices to recover qualitative features including the nonmonotonic temperature evolution of $\xi(T)$ for $\Nn=3$.
}
\label{MCversusCycle}
\end{figure*}

\section{\label{sec:compare}Comparison}
In this section we contrast the strengths and weaknesses of the Monte Carlo treatment and of the cycle expansion, starting with their computational efficiencies. Generically, given $N_{\rm TM}$ transfer/transition matrices to draw from, the number of terms to be evaluated asymptotically grows as $N_{\rm TM}^k$  at the $k$-th order of the cycle expansion. Hence, while the cycle-expansion method provides a computationally efficient method when $N_{\rm TM}$ is of order one, the attainable numerical accuracy quickly deteriorates with increasing $N_{\rm TM}$, as previously noted (see, e.g., Ref.~\onlinecite{Vanneste10}). Although a careful comparison of computational costs depends on implementation details, we empirically find that the cycle-expansion method converges much faster than the Monte Carlo algorithm for $N_{\rm TM}=2$, while its efficiency already lags for $N_{\rm TM}=4$, at least as far as the Lyapunov exponent is concerned. For the models at hand, the computational cost of the cycle expansion can be curtailed by setting $J^{[1]}>0$ through spin redefinitions, which reduces $N_{\rm TM}=2^{\Nn}$ by a factor of two. With this trick, we have carried out the cycle expansion to order $k=11$ for the NNN ($\Nn=2$) and $\Nn=3$ models, which suffices to recover Monte Carlo results within their accuracy (see Fig.~\ref{MCversusCycle}). Comparable accuracy is, however, harder to achieve for $\Nn\geq4$. As far as the attainable numerical accuracy of the Lyapunov exponent is concerned, Monte Carlo algorithm thus almost always outperforms the cycle-expansion method.

\begin{figure}[t]
\centerline{
\hspace{-0.1in}
\subfloat[]{\includegraphics[width=0.630\textwidth]{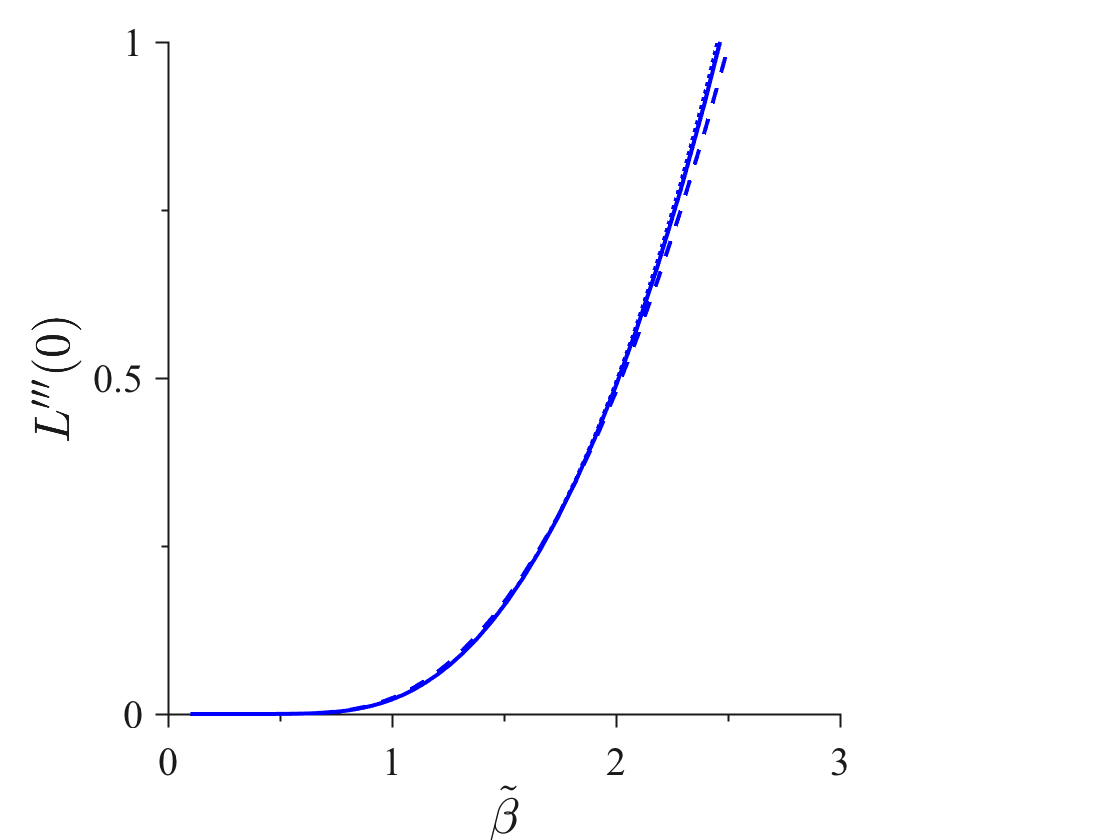}}\quad%
\hspace{-0.5in}
\subfloat[]{\includegraphics[width=0.630\textwidth]{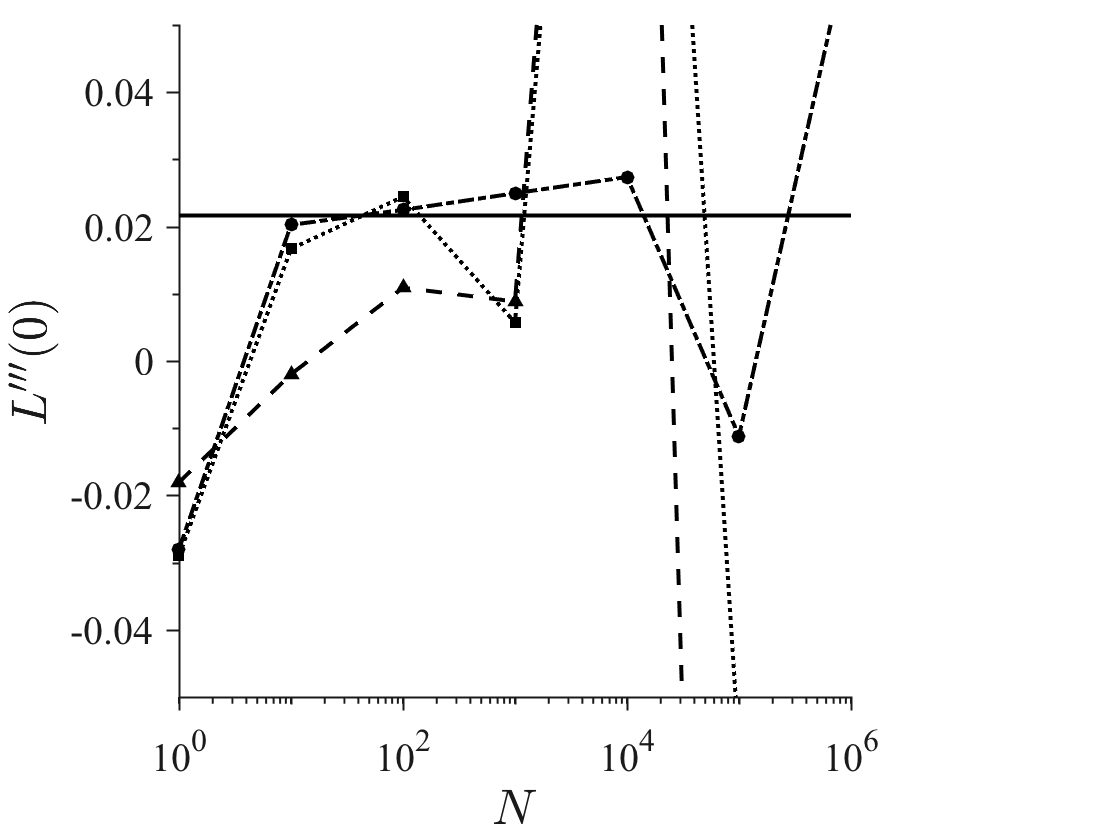}}
\hspace{-0.4in}
}
\caption{(a)  The skewness for the NNN model obtained  by the cycle expansion at order $k=3$ (dashed), 7 (dotted), 11 (dash-dotted), and 15 (solid) quickly converges with increasing $k$. (b) The skewness at $\tilde{\beta}=1$ evaluated directly through Monte Carlo simulations as a function of the number of spins $\Ns$ with a fixed number of disorder realization $\NJ=10^1$ (dashed-triangle), $10^3$ (dotted-square), and $10^5$ (dash-dotted-circle) fluctuates around the cycle expansion result for $k=15$ (solid line). The skewness and higher moments are associated with fine features of the distribution that otherwise approaches a sharply-peaked Gaussian distribution for large $\Ns$, making their reliable estimates by simple Monte Carlo numerically challenging.}
\label{skew}
\end{figure}

A different balance is, however, reached for the Lyapunov susceptibility and higher moments. With a naive implementation of the Monte Carlo algorithm, it becomes increasingly challenging to assess quantities related to the large-deviation scaling, such as skewness and kurtosis (see, however, Ref.~\onlinecite{Vanneste10} for an efficient resampling method). By contrast, cycle expansions do not encounter such difficulty and quickly converge (see Fig.~\ref{skew}). The cycle-expansion method thus offers a reliable computational tool for assessing higher moments pertaining to the large-order behavior, at least when the number of possible transfer matrices is small or when the symmetry of the problem reduces the computational cost associated with evaluating the spectral radius of cycles.

\section{Conclusion}\label{sec:conclusion}
We have developed the cycle-expansion method to compute observables pertaining to the distribution of Lyapunov exponents in systems with disorder. The cycle expansion, when its computation can be feasibly carried out to tenth order or so, reproduces the Lyapunov exponent and susceptibility results from Monte Carlo simulations, and yields far more accurate estimates of higher-order moments, such as the skewness.
The derivation of these cycle-expansion expressions, however, crucially relies on the analyticity of the generalized Lyapunov exponent. While such analyticity appears physically reasonable in the absence of replica-symmetry-breaking, a formal proof is still lacking.
It would also be interesting to develop a method that could capture the large-order behavior of higher-dimensional systems and systems with continuous distributions of quenched randomness, for which intricate dynamical effects, such as glassiness, are expected.

\begin{acknowledgments}
We thank Mengke Lian, Jonathan C.~Mattingly, and Francesco Zamponi for stimulating discussions and suggestions.
We also acknowledge the Duke Compute Cluster for computational times.
This work was supported by a grant from the Simons Foundation (\#454937, Patrick Charbonneau).
Data relevant to this work have been archived and can be accessed at http://dx.doi.org/10.7924/G88050N6.
\end{acknowledgments}

\newpage
\appendix

\section{Properties of the generalized Lyapunov exponent}
\label{sec:prop_gen_lya}
Though the results contained in this appendix are likely known to experts in the field, we here provide their succinct derivations.

In order to prove the existence of the limit defining the generalized Lyapunov exponent, we observe that
\begin{align*}
(N+M) & L_{N+M}(n) = \ln \mathbb{E} \left[\left\Vert \prod_{i=1}^{N+M}T_{i}\right\Vert ^{n}\right]\\
 & \leq\ln \mathbb{E}  \left[\left(\left\Vert \prod_{i=1}^{N}T_{i}\right\Vert ^{n}\left\Vert \prod_{i=N+1}^{N+M}T_{i}\right\Vert ^{n}\right)\right]\\
 & \leq\ln \mathbb{E}  \left[\left\Vert \prod_{i=1}^{N}T_{i}\right\Vert ^{n}\right]+\ln \mathbb{E}  \left[\left\Vert \prod_{i=N+1}^{N+M}T_{i}\right\Vert ^{n}\right]\\
 & = N L_{N}(n)+ M L_{M}(n)\, .
\end{align*}
The last two steps hold because $T_i$ is an independent and identically distributed sequence, although the argument can also be extended to handle random matrices generated by a finite-memory Markov process.
The result shows that $N L_{N}(n)$ is subadditive in $N$ and implies that
\[
L(n)\equiv\lim_{N\to\infty}\frac{1}{N}L_{N}(n)=\inf_{N\geq1}\frac{1}{N}L_{N}(n)
\]
exists for  $n\in [0,\infty)$. Similarly, the derivative
\[
L_{N}'(n)=\frac{\mathbb{E} \left[\left\Vert \prod_{i=1}^{N}T_{i}\right\Vert^{n} \ln\left\Vert \prod_{i=1}^{N}T_{i}\right\Vert \right]}{\mathbb{E} \left[\left\Vert \prod_{i=1}^{N}T_{i}\right\Vert ^{n}\right]}
\]
exists for all $n \in [0,\infty)$ and $L_N ' (0) = \mathbb{E} \left[ \ln\left\Vert \prod_{i=1}^{N}T_{i}\right\Vert \right]$.

The convexity of $L_N (n)$ can be shown as
\begin{align*}
L_N \big( & \alpha n_0 +  (1-\alpha) n_1 \big) = \frac{1}{N} \ln \mathbb{E} \left[ \left\Vert \prod_{i=1}^{N} T_{i}\right\Vert ^{\alpha n_0 + (1-\alpha) n_1} \right] \\
& \leq \frac{1}{N} \ln \left[ \left( \mathbb{E} \left\Vert \prod_{i=1}^{N}T_{i}\right\Vert ^{n_0}  \right)^{\alpha}
\left( \mathbb{E} \left\Vert \prod_{i=1}^{N}T_{i}\right\Vert ^{n_1} \right)^{1-\alpha} \right] \\
&=  \alpha L_N (n_0)  +  (1-\alpha) L_N (n_1),
\end{align*}
where H\"{o}lder's inequality is used in the second step. Because $L_{N}(n)$ is convex and differentiable, it follows that  $L(n)$ is convex and $L_{N}'(n)\to L'(n)$ at all points where $L'(n)$ exists~\cite{convex82}.
This further implies that
\[
L'(0^{+})\geq\lim_{N\to\infty}L_{N}'(0)=\lambda.
\]

\section{Notations for cycle-expansion expressions}\label{CycleNotations}
A product of $\ell$ transfer matrices, $G=T_{a_1}T_{a_2}\cdots T_{a_{\ell}}$, specifies a cycle of length $\ell(G)$. The probability of $G$ appearing amongst all the cycles of the same length $\ell$ is denoted as $p(G)$, which in our models uniformly equals $\le(\frac{1}{2^{\Nn}}\ri)^{\ell(G)}$. A cycle $G$ is prime if there is no cycle $G'$ of length $\ell(G')<\ell(G)$ with $G=(G')^{\ell(G)/\ell(G')}$~\cite{Mainieri92}. For example, $T_1 T_2 T_2 T_1$ is prime but $T_1 T_2 T_1 T_2$ is not. Prime cycles are further grouped into equivalence classes, in which two products are identified if they are related by a cyclic permutation, such as $T_1 T_2 T_4 T_5$ and $T_4 T_5 T_1 T_2$. The set of all such equivalent classes is denoted $P$. A size-$h$ subset, $\psuedo=\{G_1, G_2, \ldots, G_h\}$, is known as a pseudocycle, where $G_{\mu}\in P$ for $\mu=1,\ldots,h$ and $G_{\mu}\ne G_{\nu}$ for $\mu\ne \nu$~\cite{Nielsen97}. In particular, note that $\psuedo=\{T_1 T_2 T_3, T_2 T_3, T_1 T_2\}$ is a pseudocycle, while $\{T_1 T_2 T_3, T_1 T_2, T_1 T_2\}$ is not because the element $T_1 T_2$ is repeated. The set of all  pseudocycles, which is  the set of all subsets of equivalent classes of prime cycles, is denoted $P^{\star}$. Finally, various quantities are naturally defined as (i) the length $\ell(\psuedo)=\sum_{\mu=1}^{h}\ell(G_{\mu})$, (ii) the probability function $p(\psuedo)=\prod_{\mu=1}^{h}p(G_{\mu})$, (iii) the M{\"o}bius-function $\mathcal{M}(\psuedo)=(-1)^{h}$, and (iv) the weight $\rho(\psuedo)=\prod_{\mu=1}^{h}\rho(G_{\mu})$, where $\rho(G_{\mu})$ is the spectral radius of the matrix $G_{\mu}$.

Cycle expansions are truncated at $k$-th order by summing over all the pseudocycles of length $\ell(\psuedo)\leq k$, where the same maximum length $k$ should be used in the cycle-expansion expressions of $\partial_{z}^{s_1}\partial_{\n}^{s_2}\zeta^{-1}\le(1,0\ri)$ for all $s_1$ and $s_2$. With this truncation scheme, dilatation symmetry is preserved. That is, uniformly multiplying  transfer matrices by $c$,  $T_a\rightarrow c T_a$, makes the Lyapunov exponent $\lambda\rightarrow \lambda+\ln(c)$ while the susceptibility $\chi_{\rm L}$ and higher-moments remain invariant. To confirm this symmetry, it is useful to use the identity $\sum_{\psuedo\in P^{\star}; \ell(\psuedo)=\ell_0} \mathcal{M}(\psuedo)p(\psuedo)=\delta_{\ell_0,0}-\delta_{\ell_0,1}$ that follows from Eq.~\eqref{RuelleGod} evaluated at $\n=0$, where in particular $\zeta^{-1}(z,0)=1-z$.

\bibliography{Lyapunov_ref}

\begin{thebibliography}{65}%
\makeatletter
\providecommand \@ifxundefined [1]{%
 \@ifx{#1\undefined}
}%
\providecommand \@ifnum [1]{%
 \ifnum #1\expandafter \@firstoftwo
 \else \expandafter \@secondoftwo
 \fi
}%
\providecommand \@ifx [1]{%
 \ifx #1\expandafter \@firstoftwo
 \else \expandafter \@secondoftwo
 \fi
}%
\providecommand \natexlab [1]{#1}%
\providecommand \enquote  [1]{``#1''}%
\providecommand \bibnamefont  [1]{#1}%
\providecommand \bibfnamefont [1]{#1}%
\providecommand \citenamefont [1]{#1}%
\providecommand \href@noop [0]{\@secondoftwo}%
\providecommand \href [0]{\begingroup \@sanitize@url \@href}%
\providecommand \@href[1]{\@@startlink{#1}\@@href}%
\providecommand \@@href[1]{\endgroup#1\@@endlink}%
\providecommand \@sanitize@url [0]{\catcode `\\12\catcode `\$12\catcode
  `\&12\catcode `\#12\catcode `\^12\catcode `\_12\catcode `\%12\relax}%
\providecommand \@@startlink[1]{}%
\providecommand \@@endlink[0]{}%
\providecommand \url  [0]{\begingroup\@sanitize@url \@url }%
\providecommand \@url [1]{\endgroup\@href {#1}{\urlprefix }}%
\providecommand \urlprefix  [0]{URL }%
\providecommand \Eprint [0]{\href }%
\providecommand \doibase [0]{http://dx.doi.org/}%
\providecommand \selectlanguage [0]{\@gobble}%
\providecommand \bibinfo  [0]{\@secondoftwo}%
\providecommand \bibfield  [0]{\@secondoftwo}%
\providecommand \translation [1]{[#1]}%
\providecommand \BibitemOpen [0]{}%
\providecommand \bibitemStop [0]{}%
\providecommand \bibitemNoStop [0]{.\EOS\space}%
\providecommand \EOS [0]{\spacefactor3000\relax}%
\providecommand \BibitemShut  [1]{\csname bibitem#1\endcsname}%
\let\auto@bib@innerbib\@empty
\bibitem [{\citenamefont {Lyapunov}(1992)}]{Lyapunov}%
  \BibitemOpen
  \bibfield  {author} {\bibinfo {author} {\bibfnamefont {A.~M.}\ \bibnamefont
  {Lyapunov}},\ }\bibfield  {title} {\enquote {\bibinfo {title} {The general
  problem of the stability of motion},}\ }\href@noop {} {\bibfield  {journal}
  {\bibinfo  {journal} {International Journal of Control}\ }\textbf {\bibinfo
  {volume} {55}},\ \bibinfo {pages} {531} (\bibinfo {year} {1992})}\BibitemShut
  {NoStop}%
\bibitem [{\citenamefont {Pikovsky}\ and\ \citenamefont {Politi}(2016)}]{PP16}%
  \BibitemOpen
  \bibfield  {author} {\bibinfo {author} {\bibfnamefont {A.}~\bibnamefont
  {Pikovsky}}\ and\ \bibinfo {author} {\bibfnamefont {A.}~\bibnamefont
  {Politi}},\ }\href@noop {} {\emph {\bibinfo {title} {Lyapunov exponents: a
  tool to explore complex dynamics}}}\ (\bibinfo  {publisher} {Cambridge
  University Press},\ \bibinfo {year} {2016})\BibitemShut {NoStop}%
\bibitem [{\citenamefont {Kramers}\ and\ \citenamefont {Wannier}(1941)}]{KW41}%
  \BibitemOpen
  \bibfield  {author} {\bibinfo {author} {\bibfnamefont {H.~A.}\ \bibnamefont
  {Kramers}}\ and\ \bibinfo {author} {\bibfnamefont {G.~H.}\ \bibnamefont
  {Wannier}},\ }\bibfield  {title} {\enquote {\bibinfo {title} {Statistics of
  the two-dimensional ferromagnet. {P}art {I}},}\ }\href@noop {} {\bibfield
  {journal} {\bibinfo  {journal} {Phys. Rev.}\ }\textbf {\bibinfo {volume}
  {60}},\ \bibinfo {pages} {252} (\bibinfo {year} {1941})}\BibitemShut
  {NoStop}%
\bibitem [{\citenamefont {Blackwell}(1959)}]{Blackwell57}%
  \BibitemOpen
  \bibfield  {author} {\bibinfo {author} {\bibfnamefont {D.}~\bibnamefont
  {Blackwell}},\ }\bibfield  {title} {\enquote {\bibinfo {title} {The entropy
  of functions of finite-state {M}arkov chains},}\ }\href@noop {} {\bibfield
  {journal} {\bibinfo  {journal} {Mathematika}\ }\textbf {\bibinfo {volume}
  {3}},\ \bibinfo {pages} {143--150} (\bibinfo {year} {1959})}\BibitemShut
  {NoStop}%
\bibitem [{\citenamefont {Pfister}(2003)}]{Pfister-2003}%
  \BibitemOpen
  \bibfield  {author} {\bibinfo {author} {\bibfnamefont {Henry~D.}\
  \bibnamefont {Pfister}},\ }\emph {\bibinfo {title} {On the Capacity of Finite
  State Channels and the Analysis of Convolutional Accumulate-$m$ Codes}},\
  \href@noop {} {Ph.D. thesis},\ \bibinfo  {school} {University of California,
  San Diego}, \bibinfo {address} {CA, USA} (\bibinfo {year} {2003})\BibitemShut
  {NoStop}%
\bibitem [{\citenamefont {Holliday}\ \emph {et~al.}(2006)\citenamefont
  {Holliday}, \citenamefont {Goldsmith},\ and\ \citenamefont
  {Glynn}}]{Holliday-it06}%
  \BibitemOpen
  \bibfield  {author} {\bibinfo {author} {\bibfnamefont {T.}~\bibnamefont
  {Holliday}}, \bibinfo {author} {\bibfnamefont {A.}~\bibnamefont {Goldsmith}},
  \ and\ \bibinfo {author} {\bibfnamefont {P.}~\bibnamefont {Glynn}},\
  }\bibfield  {title} {\enquote {\bibinfo {title} {Capacity of finite state
  channels based on {L}yapunov exponents of random matrices},}\ }\href@noop {}
  {\bibfield  {journal} {\bibinfo  {journal} {IEEE Trans. Inf. Theory}\
  }\textbf {\bibinfo {volume} {52}},\ \bibinfo {pages} {3509} (\bibinfo {year}
  {2006})}\BibitemShut {NoStop}%
\bibitem [{\citenamefont {Jacquet}\ \emph {et~al.}(2008)\citenamefont
  {Jacquet}, \citenamefont {Seroussi},\ and\ \citenamefont
  {Szpankowski}}]{JSS08}%
  \BibitemOpen
  \bibfield  {author} {\bibinfo {author} {\bibfnamefont {P.}~\bibnamefont
  {Jacquet}}, \bibinfo {author} {\bibfnamefont {G.}~\bibnamefont {Seroussi}}, \
  and\ \bibinfo {author} {\bibfnamefont {W.}~\bibnamefont {Szpankowski}},\
  }\bibfield  {title} {\enquote {\bibinfo {title} {On the entropy of a hidden
  {M}arkov process},}\ }\href@noop {} {\bibfield  {journal} {\bibinfo
  {journal} {Theor. Comput. Sci.}\ }\textbf {\bibinfo {volume} {395}},\
  \bibinfo {pages} {203} (\bibinfo {year} {2008})}\BibitemShut {NoStop}%
\bibitem [{\citenamefont {Hayden}\ and\ \citenamefont {Preskill}(2007)}]{HP07}%
  \BibitemOpen
  \bibfield  {author} {\bibinfo {author} {\bibfnamefont {P.}~\bibnamefont
  {Hayden}}\ and\ \bibinfo {author} {\bibfnamefont {J.}~\bibnamefont
  {Preskill}},\ }\bibfield  {title} {\enquote {\bibinfo {title} {Black holes as
  mirrors: quantum information in random subsystems},}\ }\href@noop {}
  {\bibfield  {journal} {\bibinfo  {journal} {J. High Energy Phys.}\ }\textbf
  {\bibinfo {volume} {09}},\ \bibinfo {pages} {120} (\bibinfo {year}
  {2007})}\BibitemShut {NoStop}%
\bibitem [{\citenamefont {Sekino}\ and\ \citenamefont {Susskind}(2008)}]{SS08}%
  \BibitemOpen
  \bibfield  {author} {\bibinfo {author} {\bibfnamefont {Y.}~\bibnamefont
  {Sekino}}\ and\ \bibinfo {author} {\bibfnamefont {L.}~\bibnamefont
  {Susskind}},\ }\bibfield  {title} {\enquote {\bibinfo {title} {Fast
  scramblers},}\ }\href@noop {} {\bibfield  {journal} {\bibinfo  {journal} {J.
  High Energy Phys.}\ }\textbf {\bibinfo {volume} {10}},\ \bibinfo {pages}
  {065} (\bibinfo {year} {2008})}\BibitemShut {NoStop}%
\bibitem [{\citenamefont {Maldacena}\ \emph {et~al.}(2016)\citenamefont
  {Maldacena}, \citenamefont {Shenker},\ and\ \citenamefont
  {Stanford}}]{MSS16}%
  \BibitemOpen
  \bibfield  {author} {\bibinfo {author} {\bibfnamefont {J.}~\bibnamefont
  {Maldacena}}, \bibinfo {author} {\bibfnamefont {S.~H.}\ \bibnamefont
  {Shenker}}, \ and\ \bibinfo {author} {\bibfnamefont {D.}~\bibnamefont
  {Stanford}},\ }\bibfield  {title} {\enquote {\bibinfo {title} {A bound on
  chaos},}\ }\href@noop {} {\bibfield  {journal} {\bibinfo  {journal} {J. High
  Energy Phys.}\ }\textbf {\bibinfo {volume} {1608}},\ \bibinfo {pages} {106}
  (\bibinfo {year} {2016})}\BibitemShut {NoStop}%
\bibitem [{\citenamefont {Kurchan}(2016)}]{Kurchan16}%
  \BibitemOpen
  \bibfield  {author} {\bibinfo {author} {\bibfnamefont {J.}~\bibnamefont
  {Kurchan}},\ }\bibfield  {title} {\enquote {\bibinfo {title} {Quantum bound
  to chaos and the semiclassical limit},}\ }\href@noop {} {\  (\bibinfo {year}
  {2016})},\ \Eprint {http://arxiv.org/abs/1612.01278} {arXiv:1612.01278
  [cond-mat.stat-mech]} \BibitemShut {NoStop}%
\bibitem [{\citenamefont {Crisanti}\ \emph {et~al.}(2012)\citenamefont
  {Crisanti}, \citenamefont {Paladin},\ and\ \citenamefont {Vulpiani}}]{CPV12}%
  \BibitemOpen
  \bibfield  {author} {\bibinfo {author} {\bibfnamefont {A.}~\bibnamefont
  {Crisanti}}, \bibinfo {author} {\bibfnamefont {G.}~\bibnamefont {Paladin}}, \
  and\ \bibinfo {author} {\bibfnamefont {A.}~\bibnamefont {Vulpiani}},\
  }\href@noop {} {\emph {\bibinfo {title} {{Products of Random Matrices: in
  Statistical Physics (Vol. 104)}}}}\ (\bibinfo  {publisher} {Springer Science
  \& Business Media},\ \bibinfo {year} {2012})\BibitemShut {NoStop}%
\bibitem [{\citenamefont {M\'ezard}\ \emph {et~al.}(1987)\citenamefont
  {M\'ezard}, \citenamefont {Parisi},\ and\ \citenamefont {Virasoro}}]{MPV87}%
  \BibitemOpen
  \bibfield  {author} {\bibinfo {author} {\bibfnamefont {M.}~\bibnamefont
  {M\'ezard}}, \bibinfo {author} {\bibfnamefont {G.}~\bibnamefont {Parisi}}, \
  and\ \bibinfo {author} {\bibfnamefont {M.}~\bibnamefont {Virasoro}},\
  }\href@noop {} {\emph {\bibinfo {title} {{Spin glass theory and beyond}}}}\
  (\bibinfo  {publisher} {World Scientific},\ \bibinfo {year}
  {1987})\BibitemShut {NoStop}%
\bibitem [{\citenamefont {Nielsen}(1997)}]{Nielsen97}%
  \BibitemOpen
  \bibfield  {author} {\bibinfo {author} {\bibfnamefont {J.~L.}\ \bibnamefont
  {Nielsen}},\ }\enquote {\bibinfo {title} {Lyapunov exponent for products of
  random matrices},}\ in\ \href@noop {} {\emph {\bibinfo {booktitle} {CHAOS:
  CLASSICAL AND QUANTUM -- PROJECTS}}},\ \bibinfo {editor} {edited by\ \bibinfo
  {editor} {\bibfnamefont {P.}~\bibnamefont {Cvitanovi\'{c}}}, \bibinfo
  {editor} {\bibfnamefont {R.}~\bibnamefont {Artuso}}, \bibinfo {editor}
  {\bibfnamefont {R.}~\bibnamefont {Mainieri}}, \bibinfo {editor}
  {\bibfnamefont {G.}~\bibnamefont {Tanner}}, \bibinfo {editor} {\bibfnamefont
  {G.}~\bibnamefont {Vattay}}, \bibinfo {editor} {\bibfnamefont
  {N.}~\bibnamefont {Whelan}}, \ and\ \bibinfo {editor} {\bibfnamefont
  {A.}~\bibnamefont {Wirzba}}}\ (\bibinfo {year} {1997})\ \bibinfo {note} {at
  \url{http://chaosbook.org/projects/index.shtml}, last accessed on
  2017-06-27}\BibitemShut {NoStop}%
\bibitem [{\citenamefont {Benettin}\ \emph {et~al.}(1976)\citenamefont
  {Benettin}, \citenamefont {Galgani},\ and\ \citenamefont {Strelcyn}}]{BGS76}%
  \BibitemOpen
  \bibfield  {author} {\bibinfo {author} {\bibfnamefont {G.}~\bibnamefont
  {Benettin}}, \bibinfo {author} {\bibfnamefont {L.}~\bibnamefont {Galgani}}, \
  and\ \bibinfo {author} {\bibfnamefont {J.~M.}\ \bibnamefont {Strelcyn}},\
  }\bibfield  {title} {\enquote {\bibinfo {title} {Kolmogorov entropy and
  numerical experiments},}\ }\href@noop {} {\bibfield  {journal} {\bibinfo
  {journal} {Phys. Rev. A}\ }\textbf {\bibinfo {volume} {14}},\ \bibinfo
  {pages} {2338} (\bibinfo {year} {1976})}\BibitemShut {NoStop}%
\bibitem [{\citenamefont {Benettin}\ \emph {et~al.}(1980)\citenamefont
  {Benettin}, \citenamefont {Galgani}, \citenamefont {Giorgilli},\ and\
  \citenamefont {Strelcyn}}]{BGGS80}%
  \BibitemOpen
  \bibfield  {author} {\bibinfo {author} {\bibfnamefont {G.}~\bibnamefont
  {Benettin}}, \bibinfo {author} {\bibfnamefont {L.}~\bibnamefont {Galgani}},
  \bibinfo {author} {\bibfnamefont {A.}~\bibnamefont {Giorgilli}}, \ and\
  \bibinfo {author} {\bibfnamefont {J.~M.}\ \bibnamefont {Strelcyn}},\
  }\bibfield  {title} {\enquote {\bibinfo {title} {Lyapunov characteristic
  exponents for smooth dynamical systems and for {H}amiltonian systems; a
  method for computing all of them. {P}art 1: {T}heory},}\ }\href@noop {}
  {\bibfield  {journal} {\bibinfo  {journal} {Meccanica}\ }\textbf {\bibinfo
  {volume} {15}},\ \bibinfo {pages} {9} (\bibinfo {year} {1980})}\BibitemShut
  {NoStop}%
\bibitem [{\citenamefont {Vanneste}(2010)}]{Vanneste10}%
  \BibitemOpen
  \bibfield  {author} {\bibinfo {author} {\bibfnamefont {J.}~\bibnamefont
  {Vanneste}},\ }\bibfield  {title} {\enquote {\bibinfo {title} {Estimating
  generalized {L}yapunov exponents for products of random matrices},}\
  }\href@noop {} {\bibfield  {journal} {\bibinfo  {journal} {Phys. Rev. E}\
  }\textbf {\bibinfo {volume} {81}},\ \bibinfo {pages} {036701} (\bibinfo
  {year} {2010})}\BibitemShut {NoStop}%
\bibitem [{\citenamefont {Gardner}\ \emph {et~al.}(1984)\citenamefont
  {Gardner}, \citenamefont {J.~Itzykson},\ and\ \citenamefont
  {Derrida}}]{GID84}%
  \BibitemOpen
  \bibfield  {author} {\bibinfo {author} {\bibfnamefont {E.}~\bibnamefont
  {Gardner}}, \bibinfo {author} {\bibfnamefont {C.}~\bibnamefont
  {J.~Itzykson}}, \ and\ \bibinfo {author} {\bibfnamefont {B.}~\bibnamefont
  {Derrida}},\ }\bibfield  {title} {\enquote {\bibinfo {title} {The {L}aplacian
  on a random one-dimensional lattice},}\ }\href@noop {} {\bibfield  {journal}
  {\bibinfo  {journal} {J. Phys. A: Math. Gen.}\ }\textbf {\bibinfo {volume}
  {17}},\ \bibinfo {pages} {1093} (\bibinfo {year} {1984})}\BibitemShut
  {NoStop}%
\bibitem [{\citenamefont {Derrida}\ and\ \citenamefont {Gardner}(1984)}]{DG84}%
  \BibitemOpen
  \bibfield  {author} {\bibinfo {author} {\bibfnamefont {B.}~\bibnamefont
  {Derrida}}\ and\ \bibinfo {author} {\bibfnamefont {E.}~\bibnamefont
  {Gardner}},\ }\bibfield  {title} {\enquote {\bibinfo {title} {Lyapounov
  exponent of the one dimensional {A}nderson model : weak disorder
  expansions},}\ }\href@noop {} {\bibfield  {journal} {\bibinfo  {journal} {J.
  Phys. France}\ }\textbf {\bibinfo {volume} {45}},\ \bibinfo {pages} {1283}
  (\bibinfo {year} {1984})}\BibitemShut {NoStop}%
\bibitem [{\citenamefont {Deutsch}\ and\ \citenamefont {Paladin}(1989)}]{DP89}%
  \BibitemOpen
  \bibfield  {author} {\bibinfo {author} {\bibfnamefont {J.~M.}\ \bibnamefont
  {Deutsch}}\ and\ \bibinfo {author} {\bibfnamefont {G.}~\bibnamefont
  {Paladin}},\ }\bibfield  {title} {\enquote {\bibinfo {title} {Product of
  random matrices in a microcanonical ensemble},}\ }\href@noop {} {\bibfield
  {journal} {\bibinfo  {journal} {Phys. Rev. Lett.}\ }\textbf {\bibinfo
  {volume} {62}},\ \bibinfo {pages} {695} (\bibinfo {year} {1989})}\BibitemShut
  {NoStop}%
\bibitem [{\citenamefont {Mainieri}(1992)}]{Mainieri92}%
  \BibitemOpen
  \bibfield  {author} {\bibinfo {author} {\bibfnamefont {R.}~\bibnamefont
  {Mainieri}},\ }\bibfield  {title} {\enquote {\bibinfo {title} {Zeta function
  for the {L}yapunov exponent of a product of random matrices},}\ }\href@noop
  {} {\bibfield  {journal} {\bibinfo  {journal} {Phys. Rev. Lett.}\ }\textbf
  {\bibinfo {volume} {68}},\ \bibinfo {pages} {1965} (\bibinfo {year}
  {1992})}\BibitemShut {NoStop}%
\bibitem [{\citenamefont {Derrida}\ and\ \citenamefont
  {Hilhorst}(1983)}]{DH83}%
  \BibitemOpen
  \bibfield  {author} {\bibinfo {author} {\bibfnamefont {B.}~\bibnamefont
  {Derrida}}\ and\ \bibinfo {author} {\bibfnamefont {H.~J.}\ \bibnamefont
  {Hilhorst}},\ }\bibfield  {title} {\enquote {\bibinfo {title} {Singular
  behaviour of certain infinite products of random $2\times 2$ matrices},}\
  }\href@noop {} {\bibfield  {journal} {\bibinfo  {journal} {J. Phys. A: Math.
  Gen.}\ }\textbf {\bibinfo {volume} {16}},\ \bibinfo {pages} {2641} (\bibinfo
  {year} {1983})}\BibitemShut {NoStop}%
\bibitem [{\citenamefont {Weigt}\ and\ \citenamefont {Monasson}(1996)}]{WM96}%
  \BibitemOpen
  \bibfield  {author} {\bibinfo {author} {\bibfnamefont {M.}~\bibnamefont
  {Weigt}}\ and\ \bibinfo {author} {\bibfnamefont {R.}~\bibnamefont
  {Monasson}},\ }\bibfield  {title} {\enquote {\bibinfo {title} {Replica
  structure of one-dimensional disordered ising models},}\ }\href@noop {}
  {\bibfield  {journal} {\bibinfo  {journal} {Europhys. Lett.}\ }\textbf
  {\bibinfo {volume} {36}},\ \bibinfo {pages} {209} (\bibinfo {year}
  {1996})}\BibitemShut {NoStop}%
\bibitem [{\citenamefont {Paladin}\ and\ \citenamefont {Serva}(1992)}]{PS92}%
  \BibitemOpen
  \bibfield  {author} {\bibinfo {author} {\bibfnamefont {G.}~\bibnamefont
  {Paladin}}\ and\ \bibinfo {author} {\bibfnamefont {M.}~\bibnamefont
  {Serva}},\ }\bibfield  {title} {\enquote {\bibinfo {title} {Analytic solution
  of the random {I}sing model in one dimension},}\ }\href@noop {} {\bibfield
  {journal} {\bibinfo  {journal} {Phys. Rev. Lett.}\ }\textbf {\bibinfo
  {volume} {69}},\ \bibinfo {pages} {706} (\bibinfo {year} {1992})}\BibitemShut
  {NoStop}%
\bibitem [{\citenamefont {Davids}(1994)}]{Davids94}%
  \BibitemOpen
  \bibfield  {author} {\bibinfo {author} {\bibfnamefont {P.~S.}\ \bibnamefont
  {Davids}},\ }\bibfield  {title} {\enquote {\bibinfo {title} {Analytic
  structure of the one-dimensional random-bond {I}sing model},}\ }\href@noop {}
  {\bibfield  {journal} {\bibinfo  {journal} {J. Phys. A: Math. Gen.}\ }\textbf
  {\bibinfo {volume} {27}},\ \bibinfo {pages} {6703} (\bibinfo {year}
  {1994})}\BibitemShut {NoStop}%
\bibitem [{\citenamefont {Bai}(2007)}]{Bai07}%
  \BibitemOpen
  \bibfield  {author} {\bibinfo {author} {\bibfnamefont {Z.-Q.}\ \bibnamefont
  {Bai}},\ }\bibfield  {title} {\enquote {\bibinfo {title} {On the cycle
  expansion for the {L}yapunov exponent of a product of random matrices},}\
  }\href@noop {} {\bibfield  {journal} {\bibinfo  {journal} {J. Phys. A: Math.
  Theor.}\ }\textbf {\bibinfo {volume} {40}},\ \bibinfo {pages} {8315}
  (\bibinfo {year} {2007})}\BibitemShut {NoStop}%
\bibitem [{\citenamefont {Bai}(2009)}]{Bai09}%
  \BibitemOpen
  \bibfield  {author} {\bibinfo {author} {\bibfnamefont {Z.-Q.}\ \bibnamefont
  {Bai}},\ }\bibfield  {title} {\enquote {\bibinfo {title} {An infinite
  transfer matrix approach to the product of random $2\times 2$ positive
  matrices},}\ }\href@noop {} {\bibfield  {journal} {\bibinfo  {journal} {J.
  Phys. A: Math. Theor.}\ }\textbf {\bibinfo {volume} {42}},\ \bibinfo {pages}
  {015003} (\bibinfo {year} {2009})}\BibitemShut {NoStop}%
\bibitem [{\citenamefont {Fujisaka}(1983)}]{Fujisaka83}%
  \BibitemOpen
  \bibfield  {author} {\bibinfo {author} {\bibfnamefont {H.}~\bibnamefont
  {Fujisaka}},\ }\bibfield  {title} {\enquote {\bibinfo {title} {Statistical
  dynamics generated by fluctuations of local {L}yapunov exponents},}\
  }\href@noop {} {\bibfield  {journal} {\bibinfo  {journal} {Prog. Theor.
  Phys.}\ }\textbf {\bibinfo {volume} {70}},\ \bibinfo {pages} {1264} (\bibinfo
  {year} {1983})}\BibitemShut {NoStop}%
\bibitem [{\citenamefont {Ishitani}(1977)}]{Ishitani77}%
  \BibitemOpen
  \bibfield  {author} {\bibinfo {author} {\bibfnamefont {H.}~\bibnamefont
  {Ishitani}},\ }\bibfield  {title} {\enquote {\bibinfo {title} {A central
  limit theorem for the subadditive process and its application to products of
  random matrices},}\ }\href@noop {} {\bibfield  {journal} {\bibinfo  {journal}
  {Publ. Res. Inst. Math. Sci.}\ }\textbf {\bibinfo {volume} {12}},\ \bibinfo
  {pages} {565} (\bibinfo {year} {1977})}\BibitemShut {NoStop}%
\bibitem [{\citenamefont {Aizenman}\ \emph {et~al.}(1987)\citenamefont
  {Aizenman}, \citenamefont {Lebowitz},\ and\ \citenamefont {Ruelle}}]{ALR87}%
  \BibitemOpen
  \bibfield  {author} {\bibinfo {author} {\bibfnamefont {M.}~\bibnamefont
  {Aizenman}}, \bibinfo {author} {\bibfnamefont {J.~L.}\ \bibnamefont
  {Lebowitz}}, \ and\ \bibinfo {author} {\bibfnamefont {D.}~\bibnamefont
  {Ruelle}},\ }\bibfield  {title} {\enquote {\bibinfo {title} {Some rigorous
  results on the {S}herrington-{K}irkpatrick spin glass model},}\ }\href@noop
  {} {\bibfield  {journal} {\bibinfo  {journal} {Comm. Math. Phys.}\ }\textbf
  {\bibinfo {volume} {112}},\ \bibinfo {pages} {3} (\bibinfo {year}
  {1987})}\BibitemShut {NoStop}%
\bibitem [{\citenamefont {Baik}\ and\ \citenamefont {Lee}(2016)}]{BL16}%
  \BibitemOpen
  \bibfield  {author} {\bibinfo {author} {\bibfnamefont {J.}~\bibnamefont
  {Baik}}\ and\ \bibinfo {author} {\bibfnamefont {J.~O.}\ \bibnamefont {Lee}},\
  }\bibfield  {title} {\enquote {\bibinfo {title} {Fluctuations of the free
  energy of the spherical {S}herrington-{K}irkpatrick model},}\ }\href@noop {}
  {\bibfield  {journal} {\bibinfo  {journal} {J. Stat. Phys.}\ }\textbf
  {\bibinfo {volume} {165}},\ \bibinfo {pages} {185} (\bibinfo {year}
  {2016})}\BibitemShut {NoStop}%
\bibitem [{\citenamefont {Bray}\ and\ \citenamefont {Moore}(1987)}]{BM87}%
  \BibitemOpen
  \bibfield  {author} {\bibinfo {author} {\bibfnamefont {A.~J.}\ \bibnamefont
  {Bray}}\ and\ \bibinfo {author} {\bibfnamefont {M.~A.}\ \bibnamefont
  {Moore}},\ }\bibfield  {title} {\enquote {\bibinfo {title} {Chaotic nature of
  the spin-glass phase},}\ }\href@noop {} {\bibfield  {journal} {\bibinfo
  {journal} {Phys. Rev. Lett.}\ }\textbf {\bibinfo {volume} {58}},\ \bibinfo
  {pages} {57} (\bibinfo {year} {1987})}\BibitemShut {NoStop}%
\bibitem [{\citenamefont {Bouchaud}\ \emph {et~al.}(2003)\citenamefont
  {Bouchaud}, \citenamefont {Krzakala},\ and\ \citenamefont {Martin}}]{BKM03}%
  \BibitemOpen
  \bibfield  {author} {\bibinfo {author} {\bibfnamefont {J.-P.}\ \bibnamefont
  {Bouchaud}}, \bibinfo {author} {\bibfnamefont {F.}~\bibnamefont {Krzakala}},
  \ and\ \bibinfo {author} {\bibfnamefont {O.~C.}\ \bibnamefont {Martin}},\
  }\bibfield  {title} {\enquote {\bibinfo {title} {Energy exponents and
  corrections to scaling in {I}sing spin glasses},}\ }\href@noop {} {\bibfield
  {journal} {\bibinfo  {journal} {Phys. Rev. B}\ }\textbf {\bibinfo {volume}
  {68}},\ \bibinfo {pages} {224404} (\bibinfo {year} {2003})}\BibitemShut
  {NoStop}%
\bibitem [{\citenamefont {Aspelmeier}(2008)}]{Aspelmeier08}%
  \BibitemOpen
  \bibfield  {author} {\bibinfo {author} {\bibfnamefont {T.}~\bibnamefont
  {Aspelmeier}},\ }\bibfield  {title} {\enquote {\bibinfo {title} {Free-energy
  fluctuations and chaos in the {S}herrington-{K}irkpatrick model},}\
  }\href@noop {} {\bibfield  {journal} {\bibinfo  {journal} {Phys. Rev. Lett.}\
  }\textbf {\bibinfo {volume} {100}},\ \bibinfo {pages} {117205} (\bibinfo
  {year} {2008})}\BibitemShut {NoStop}%
\bibitem [{\citenamefont {Benzi}\ \emph {et~al.}(1985)\citenamefont {Benzi},
  \citenamefont {Paladin}, \citenamefont {Parisi},\ and\ \citenamefont
  {Vulpiani}}]{BPPV85}%
  \BibitemOpen
  \bibfield  {author} {\bibinfo {author} {\bibfnamefont {R}~\bibnamefont
  {Benzi}}, \bibinfo {author} {\bibfnamefont {G}~\bibnamefont {Paladin}},
  \bibinfo {author} {\bibfnamefont {G}~\bibnamefont {Parisi}}, \ and\ \bibinfo
  {author} {\bibfnamefont {A}~\bibnamefont {Vulpiani}},\ }\bibfield  {title}
  {\enquote {\bibinfo {title} {Characterisation of intermittency in chaotic
  systems},}\ }\href@noop {} {\bibfield  {journal} {\bibinfo  {journal} {J.
  Phys. A: Math. Gen.}\ }\textbf {\bibinfo {volume} {18}},\ \bibinfo {pages}
  {2157} (\bibinfo {year} {1985})}\BibitemShut {NoStop}%
\bibitem [{\citenamefont {Frisch}(1995)}]{Frisch95}%
  \BibitemOpen
  \bibfield  {author} {\bibinfo {author} {\bibfnamefont {U.}~\bibnamefont
  {Frisch}},\ }\href@noop {} {\emph {\bibinfo {title} {Turbulence: The Legacy
  of {A}. {N}. {K}olmogorov}}}\ (\bibinfo  {publisher} {Cambridge University
  Press},\ \bibinfo {year} {1995})\BibitemShut {NoStop}%
\bibitem [{\citenamefont {Cecconi}\ \emph {et~al.}(2014)\citenamefont
  {Cecconi}, \citenamefont {Cencini}, \citenamefont {Puglisi}, \citenamefont
  {Vergni},\ and\ \citenamefont {Vulpiani}}]{CCPVV14}%
  \BibitemOpen
  \bibfield  {author} {\bibinfo {author} {\bibfnamefont {F.}~\bibnamefont
  {Cecconi}}, \bibinfo {author} {\bibfnamefont {M.}~\bibnamefont {Cencini}},
  \bibinfo {author} {\bibfnamefont {A.}~\bibnamefont {Puglisi}}, \bibinfo
  {author} {\bibfnamefont {D.}~\bibnamefont {Vergni}}, \ and\ \bibinfo {author}
  {\bibfnamefont {A.}~\bibnamefont {Vulpiani}},\ }\enquote {\bibinfo {title}
  {From the law of large numbers to large deviation theory in statistical
  physics: An introduction},}\ in\ \href@noop {} {\emph {\bibinfo {booktitle}
  {Large Deviations in Physics: The Legacy of the Law of Large Numbers}}},\
  \bibinfo {editor} {edited by\ \bibinfo {editor} {\bibfnamefont
  {A.}~\bibnamefont {Vulpiani}}, \bibinfo {editor} {\bibfnamefont
  {F.}~\bibnamefont {Cecconi}}, \bibinfo {editor} {\bibfnamefont
  {M.}~\bibnamefont {Cencini}}, \bibinfo {editor} {\bibfnamefont
  {A.}~\bibnamefont {Puglisi}}, \ and\ \bibinfo {editor} {\bibfnamefont
  {D.}~\bibnamefont {Vergni}}}\ (\bibinfo  {publisher} {Springer Berlin
  Heidelberg},\ \bibinfo {address} {Berlin, Heidelberg},\ \bibinfo {year}
  {2014})\ pp.\ \bibinfo {pages} {1--27}\BibitemShut {NoStop}%
\bibitem [{\citenamefont {Cvitanovi\'{c}}(1988)}]{Cvitanovic88}%
  \BibitemOpen
  \bibfield  {author} {\bibinfo {author} {\bibfnamefont {P.}~\bibnamefont
  {Cvitanovi\'{c}}},\ }\bibfield  {title} {\enquote {\bibinfo {title}
  {Invariant measurement of strange sets in terms of cycles},}\ }\href@noop {}
  {\bibfield  {journal} {\bibinfo  {journal} {Phys. Rev. Lett.}\ }\textbf
  {\bibinfo {volume} {61}},\ \bibinfo {pages} {2729} (\bibinfo {year}
  {1988})}\BibitemShut {NoStop}%
\bibitem [{\citenamefont {Artuso}\ \emph {et~al.}(1990)\citenamefont {Artuso},
  \citenamefont {Aurell},\ and\ \citenamefont {Cvitanovi\'{c}}}]{AAC90}%
  \BibitemOpen
  \bibfield  {author} {\bibinfo {author} {\bibfnamefont {R.}~\bibnamefont
  {Artuso}}, \bibinfo {author} {\bibfnamefont {E.}~\bibnamefont {Aurell}}, \
  and\ \bibinfo {author} {\bibfnamefont {P.}~\bibnamefont {Cvitanovi\'{c}}},\
  }\bibfield  {title} {\enquote {\bibinfo {title} {Recycling of strange sets:
  {I}. {C}ycle expansions},}\ }\href@noop {} {\bibfield  {journal} {\bibinfo
  {journal} {Nonlinearity}\ }\textbf {\bibinfo {volume} {3}},\ \bibinfo {pages}
  {325} (\bibinfo {year} {1990})}\BibitemShut {NoStop}%
\bibitem [{\citenamefont {Gelfand}(1941)}]{Gelfand41}%
  \BibitemOpen
  \bibfield  {author} {\bibinfo {author} {\bibfnamefont {I.}~\bibnamefont
  {Gelfand}},\ }\bibfield  {title} {\enquote {\bibinfo {title} {Zur {T}heorie
  der {C}haraktere der {A}belschen topologischen {G}ruppen},}\ }\href@noop {}
  {\bibfield  {journal} {\bibinfo  {journal} {Rec. Math. [Mat. Sbornik] N.S.}\
  }\textbf {\bibinfo {volume} {9}},\ \bibinfo {pages} {49} (\bibinfo {year}
  {1941})}\BibitemShut {NoStop}%
\bibitem [{\citenamefont {Selke}\ and\ \citenamefont {Fisher}(1979)}]{SF79}%
  \BibitemOpen
  \bibfield  {author} {\bibinfo {author} {\bibfnamefont {W.}~\bibnamefont
  {Selke}}\ and\ \bibinfo {author} {\bibfnamefont {M.~E.}\ \bibnamefont
  {Fisher}},\ }\bibfield  {title} {\enquote {\bibinfo {title} {{M}onte {C}arlo
  study of the spatially modulated phase in an ising model},}\ }\href@noop {}
  {\bibfield  {journal} {\bibinfo  {journal} {Phys. Rev. B}\ }\textbf {\bibinfo
  {volume} {20}},\ \bibinfo {pages} {257} (\bibinfo {year} {1979})}\BibitemShut
  {NoStop}%
\bibitem [{foo({\natexlab{a}})}]{footnote_Ishitani}%
  \BibitemOpen
  \href@noop \ \bibinfo {note} {The theorem in
  Ref.~\cite{Ishitani77} stipulates that the entries of transfer matrices must
  all be positive while each transfer matrix in our generalized nearest
  neighbor models has zeros for $\Nn\geq2$. In order to apply the theorem, we
  can instead look at products of $\Nn$ such matrices, which have all positive
  entries and are also independently and identically distributed.}\BibitemShut
  {Stop}%
\bibitem [{\citenamefont {Sherrington}\ and\ \citenamefont
  {Kirkpatrick}(1975)}]{SK75}%
  \BibitemOpen
  \bibfield  {author} {\bibinfo {author} {\bibfnamefont {D.}~\bibnamefont
  {Sherrington}}\ and\ \bibinfo {author} {\bibfnamefont {S.}~\bibnamefont
  {Kirkpatrick}},\ }\bibfield  {title} {\enquote {\bibinfo {title} {Solvable
  model of a spin-glass},}\ }\href@noop {} {\bibfield  {journal} {\bibinfo
  {journal} {Phys. Rev. Lett.}\ }\textbf {\bibinfo {volume} {35}},\ \bibinfo
  {pages} {1792} (\bibinfo {year} {1975})}\BibitemShut {NoStop}%
\bibitem [{\citenamefont {Kotliar}\ \emph {et~al.}(1983)\citenamefont
  {Kotliar}, \citenamefont {Anderson},\ and\ \citenamefont {Stein}}]{KAS83}%
  \BibitemOpen
  \bibfield  {author} {\bibinfo {author} {\bibfnamefont {G.}~\bibnamefont
  {Kotliar}}, \bibinfo {author} {\bibfnamefont {P.~W.}\ \bibnamefont
  {Anderson}}, \ and\ \bibinfo {author} {\bibfnamefont {D.~L.}\ \bibnamefont
  {Stein}},\ }\bibfield  {title} {\enquote {\bibinfo {title} {One-dimensional
  spin-glass model with long-range random interactions},}\ }\href@noop {}
  {\bibfield  {journal} {\bibinfo  {journal} {Phys. Rev. B}\ }\textbf {\bibinfo
  {volume} {27}},\ \bibinfo {pages} {602(R)} (\bibinfo {year}
  {1983})}\BibitemShut {NoStop}%
\bibitem [{\citenamefont {Aspelmeier}\ \emph {et~al.}(2016)\citenamefont
  {Aspelmeier}, \citenamefont {Wang}, \citenamefont {Moore},\ and\
  \citenamefont {Katzgraber}}]{AWMK16}%
  \BibitemOpen
  \bibfield  {author} {\bibinfo {author} {\bibfnamefont {T.}~\bibnamefont
  {Aspelmeier}}, \bibinfo {author} {\bibfnamefont {W.}~\bibnamefont {Wang}},
  \bibinfo {author} {\bibfnamefont {M.~A.}\ \bibnamefont {Moore}}, \ and\
  \bibinfo {author} {\bibfnamefont {H.~G.}\ \bibnamefont {Katzgraber}},\
  }\bibfield  {title} {\enquote {\bibinfo {title} {Interface free-energy
  exponent in the one-dimensional ising spin glass with long-range interactions
  in both the droplet and broken replica symmetry regions},}\ }\href@noop {}
  {\bibfield  {journal} {\bibinfo  {journal} {Phys. Rev. E}\ }\textbf {\bibinfo
  {volume} {94}},\ \bibinfo {pages} {022116} (\bibinfo {year}
  {2016})}\BibitemShut {NoStop}%
\bibitem [{\citenamefont {Franz}\ \emph {et~al.}(2009)\citenamefont {Franz},
  \citenamefont {J\"{o}rg},\ and\ \citenamefont {Parisi}}]{FJP09}%
  \BibitemOpen
  \bibfield  {author} {\bibinfo {author} {\bibfnamefont {S.}~\bibnamefont
  {Franz}}, \bibinfo {author} {\bibfnamefont {T.}~\bibnamefont {J\"{o}rg}}, \
  and\ \bibinfo {author} {\bibfnamefont {G.}~\bibnamefont {Parisi}},\
  }\bibfield  {title} {\enquote {\bibinfo {title} {Overlap interfaces in
  hierarchical spin-glass models},}\ }\href@noop {} {\bibfield  {journal}
  {\bibinfo  {journal} {J. Stat. Mech.}\ }\textbf {\bibinfo {volume} {2009}},\
  \bibinfo {pages} {P020021} (\bibinfo {year} {2009})}\BibitemShut {NoStop}%
\bibitem [{\citenamefont {Castellana}\ and\ \citenamefont
  {Parisi}(2010)}]{CP10}%
  \BibitemOpen
  \bibfield  {author} {\bibinfo {author} {\bibfnamefont {M.}~\bibnamefont
  {Castellana}}\ and\ \bibinfo {author} {\bibfnamefont {G.}~\bibnamefont
  {Parisi}},\ }\bibfield  {title} {\enquote {\bibinfo {title} {Renormalization
  group computation of the critical exponents of hierarchical spin glasses},}\
  }\href@noop {} {\bibfield  {journal} {\bibinfo  {journal} {Phys. Rev. E}\
  }\textbf {\bibinfo {volume} {82}},\ \bibinfo {pages} {040105(R)} (\bibinfo
  {year} {2010})}\BibitemShut {NoStop}%
\bibitem [{\citenamefont {Charbonneau}\ and\ \citenamefont
  {Yaida}(2017)}]{CY17}%
  \BibitemOpen
  \bibfield  {author} {\bibinfo {author} {\bibfnamefont {P.}~\bibnamefont
  {Charbonneau}}\ and\ \bibinfo {author} {\bibfnamefont {S.}~\bibnamefont
  {Yaida}},\ }\bibfield  {title} {\enquote {\bibinfo {title} {Nontrivial
  critical fixed point for replica-symmetry-breaking transitions},}\
  }\href@noop {} {\bibfield  {journal} {\bibinfo  {journal} {Phys. Rev. Lett.}\
  }\textbf {\bibinfo {volume} {118}},\ \bibinfo {pages} {215701} (\bibinfo
  {year} {2017})}\BibitemShut {NoStop}%
\bibitem [{\citenamefont {Skokos}(2010)}]{Skokos10}%
  \BibitemOpen
  \bibfield  {author} {\bibinfo {author} {\bibfnamefont {Ch.}\ \bibnamefont
  {Skokos}},\ }\bibfield  {title} {\enquote {\bibinfo {title} {The {L}yapunov
  characteristic exponents and their computation},}\ }\href@noop {} {\bibfield
  {journal} {\bibinfo  {journal} {Lect. Notes Phys.}\ }\textbf {\bibinfo
  {volume} {790}},\ \bibinfo {pages} {63} (\bibinfo {year} {2010})}\BibitemShut
  {NoStop}%
\bibitem [{\citenamefont {Baxter}(1971)}]{Baxter71}%
  \BibitemOpen
  \bibfield  {author} {\bibinfo {author} {\bibfnamefont {R.~J.}\ \bibnamefont
  {Baxter}},\ }\bibfield  {title} {\enquote {\bibinfo {title} {Eight-vertex
  model in lattice statistics},}\ }\href@noop {} {\bibfield  {journal}
  {\bibinfo  {journal} {Phys. Rev. Lett.}\ }\textbf {\bibinfo {volume} {26}},\
  \bibinfo {pages} {832} (\bibinfo {year} {1971})}\BibitemShut {NoStop}%
\bibitem [{\citenamefont {Baxter}\ and\ \citenamefont {Wu}(1974)}]{BW74}%
  \BibitemOpen
  \bibfield  {author} {\bibinfo {author} {\bibfnamefont {R.~J.}\ \bibnamefont
  {Baxter}}\ and\ \bibinfo {author} {\bibfnamefont {F.~Y.}\ \bibnamefont
  {Wu}},\ }\bibfield  {title} {\enquote {\bibinfo {title} {Ising model on a
  triangular lattice with three-spin interactions. {I}. {T}he eigenvalue
  equation},}\ }\href@noop {} {\bibfield  {journal} {\bibinfo  {journal} {Aust.
  J. Phys.}\ }\textbf {\bibinfo {volume} {27}},\ \bibinfo {pages} {357}
  (\bibinfo {year} {1974})}\BibitemShut {NoStop}%
\bibitem [{\citenamefont {Krinsky}\ and\ \citenamefont {Mukamel}(1977)}]{KM77}%
  \BibitemOpen
  \bibfield  {author} {\bibinfo {author} {\bibfnamefont {S.}~\bibnamefont
  {Krinsky}}\ and\ \bibinfo {author} {\bibfnamefont {D.}~\bibnamefont
  {Mukamel}},\ }\bibfield  {title} {\enquote {\bibinfo {title} {Ising models
  with $n\geq2$-component order parameters},}\ }\href@noop {} {\bibfield
  {journal} {\bibinfo  {journal} {Phys. Rev. B}\ }\textbf {\bibinfo {volume}
  {16}},\ \bibinfo {pages} {2313} (\bibinfo {year} {1977})}\BibitemShut
  {NoStop}%
\bibitem [{\citenamefont {Barber}(1979)}]{Barber79}%
  \BibitemOpen
  \bibfield  {author} {\bibinfo {author} {\bibfnamefont {M.~N.}\ \bibnamefont
  {Barber}},\ }\bibfield  {title} {\enquote {\bibinfo {title} {Non-universality
  in the ising model with nearest and next-nearest neighbour interactions},}\
  }\href@noop {} {\bibfield  {journal} {\bibinfo  {journal} {J. Phys. A: Math.
  Gen.}\ }\textbf {\bibinfo {volume} {12}},\ \bibinfo {pages} {679} (\bibinfo
  {year} {1979})}\BibitemShut {NoStop}%
\bibitem [{\citenamefont {Hardy}\ \emph {et~al.}(1934)\citenamefont {Hardy},
  \citenamefont {Littlewood},\ and\ \citenamefont {P\'olya}}]{HLP34}%
  \BibitemOpen
  \bibfield  {author} {\bibinfo {author} {\bibfnamefont {G.~H.}\ \bibnamefont
  {Hardy}}, \bibinfo {author} {\bibfnamefont {J.~E.}\ \bibnamefont
  {Littlewood}}, \ and\ \bibinfo {author} {\bibfnamefont {G.}~\bibnamefont
  {P\'olya}},\ }\href@noop {} {\emph {\bibinfo {title} {{Inequalities}}}}\
  (\bibinfo  {publisher} {Cambridge University Press},\ \bibinfo {year}
  {1934})\BibitemShut {NoStop}%
\bibitem [{\citenamefont {Edwards}\ and\ \citenamefont
  {Anderson}(1975)}]{EA75}%
  \BibitemOpen
  \bibfield  {author} {\bibinfo {author} {\bibfnamefont {S.~F.}\ \bibnamefont
  {Edwards}}\ and\ \bibinfo {author} {\bibfnamefont {P.~W.}\ \bibnamefont
  {Anderson}},\ }\bibfield  {title} {\enquote {\bibinfo {title} {Theory of spin
  glasses},}\ }\href@noop {} {\bibfield  {journal} {\bibinfo  {journal} {J.
  Phys. F: Met. Phys.}\ }\textbf {\bibinfo {volume} {5}},\ \bibinfo {pages}
  {965} (\bibinfo {year} {1975})}\BibitemShut {NoStop}%
\bibitem [{\citenamefont {Pendry}(1982)}]{Pendry82}%
  \BibitemOpen
  \bibfield  {author} {\bibinfo {author} {\bibfnamefont {J.~B.}\ \bibnamefont
  {Pendry}},\ }\bibfield  {title} {\enquote {\bibinfo {title} {1{D}
  localisation and the symmetric group},}\ }\href@noop {} {\bibfield  {journal}
  {\bibinfo  {journal} {J. Phys. C}\ }\textbf {\bibinfo {volume} {15}},\
  \bibinfo {pages} {4821} (\bibinfo {year} {1982})}\BibitemShut {NoStop}%
\bibitem [{\citenamefont {Bouchard}\ \emph {et~al.}(1986)\citenamefont
  {Bouchard}, \citenamefont {Georges}, \citenamefont {Hansel}, \citenamefont
  {Le~Doussal},\ and\ \citenamefont {Maillard}}]{BGHLM86}%
  \BibitemOpen
  \bibfield  {author} {\bibinfo {author} {\bibfnamefont {J.-P.}\ \bibnamefont
  {Bouchard}}, \bibinfo {author} {\bibfnamefont {A.}~\bibnamefont {Georges}},
  \bibinfo {author} {\bibfnamefont {D.}~\bibnamefont {Hansel}}, \bibinfo
  {author} {\bibfnamefont {P.}~\bibnamefont {Le~Doussal}}, \ and\ \bibinfo
  {author} {\bibfnamefont {J.~M.}\ \bibnamefont {Maillard}},\ }\bibfield
  {title} {\enquote {\bibinfo {title} {Rigorous bounds and the replica method
  for products of random matrices},}\ }\href@noop {} {\bibfield  {journal}
  {\bibinfo  {journal} {J. Phys. A: Math. Gen.}\ }\textbf {\bibinfo {volume}
  {19}},\ \bibinfo {pages} {L1145} (\bibinfo {year} {1986})}\BibitemShut
  {NoStop}%
\bibitem [{\citenamefont {de~Oliveira}\ and\ \citenamefont
  {Petri}(1996)}]{OP96}%
  \BibitemOpen
  \bibfield  {author} {\bibinfo {author} {\bibfnamefont {M.~J.}\ \bibnamefont
  {de~Oliveira}}\ and\ \bibinfo {author} {\bibfnamefont {A.}~\bibnamefont
  {Petri}},\ }\bibfield  {title} {\enquote {\bibinfo {title} {Generalized
  {L}yapunov exponents for products of correlated random matrices},}\
  }\href@noop {} {\bibfield  {journal} {\bibinfo  {journal} {Phys. Rev. E}\
  }\textbf {\bibinfo {volume} {53}},\ \bibinfo {pages} {2960} (\bibinfo {year}
  {1996})}\BibitemShut {NoStop}%
\bibitem [{\citenamefont {van Hemmen}\ and\ \citenamefont
  {Palmer}(1979)}]{HP79}%
  \BibitemOpen
  \bibfield  {author} {\bibinfo {author} {\bibfnamefont {J.~L.}\ \bibnamefont
  {van Hemmen}}\ and\ \bibinfo {author} {\bibfnamefont {R.~G.}\ \bibnamefont
  {Palmer}},\ }\bibfield  {title} {\enquote {\bibinfo {title} {The replica
  method and solvable spin glass model},}\ }\href@noop {} {\bibfield  {journal}
  {\bibinfo  {journal} {J. Phys. A: Math. Gen.}\ }\textbf {\bibinfo {volume}
  {12}},\ \bibinfo {pages} {563} (\bibinfo {year} {1979})}\BibitemShut
  {NoStop}%
\bibitem [{foo({\natexlab{b}})}]{footnote_large}%
  \BibitemOpen
  \href@noop \ \bibinfo {note} {Note that in the
  dynamical system literature the function $L(\n)$ is known as the generalized
  Lyapunov exponent~\cite{Fujisaka83,BPPV85}, which is related through a
  Legendre transform to the Cram\'{e}r function governing the large-deviation
  behavior of the Lyapunov exponent
  $\lambda_{\Ns}$~\cite{CCPVV14}.}\BibitemShut {Stop}%
\bibitem [{\citenamefont {Ruelle}(1979)}]{Ruelle79}%
  \BibitemOpen
  \bibfield  {author} {\bibinfo {author} {\bibfnamefont {D.}~\bibnamefont
  {Ruelle}},\ }\bibfield  {title} {\enquote {\bibinfo {title} {Analycity
  properties of the characteristic exponents of random matrix products},}\
  }\href@noop {} {\bibfield  {journal} {\bibinfo  {journal} {Adv. Math.}\
  }\textbf {\bibinfo {volume} {32}},\ \bibinfo {pages} {68} (\bibinfo {year}
  {1979})}\BibitemShut {NoStop}%
\bibitem [{\citenamefont {Ruelle}(1978)}]{Ruelle78}%
  \BibitemOpen
  \bibfield  {author} {\bibinfo {author} {\bibfnamefont {D.}~\bibnamefont
  {Ruelle}},\ }\href@noop {} {\emph {\bibinfo {title} {{Thermodynamic
  formalism}}}}\ (\bibinfo  {publisher} {Addison-Wesley},\ \bibinfo {year}
  {1978})\BibitemShut {NoStop}%
\bibitem [{\citenamefont {Ruelle}(2002)}]{Ruelle02}%
  \BibitemOpen
  \bibfield  {author} {\bibinfo {author} {\bibfnamefont {D.}~\bibnamefont
  {Ruelle}},\ }\bibfield  {title} {\enquote {\bibinfo {title} {Dynamical zeta
  functions and transfer operators},}\ }\href@noop {} {\bibfield  {journal}
  {\bibinfo  {journal} {Notices Amer. Math. Soc.}\ }\textbf {\bibinfo {volume}
  {49}},\ \bibinfo {pages} {887} (\bibinfo {year} {2002})}\BibitemShut
  {NoStop}%
\bibitem [{foo({\natexlab{c}})}]{footnote_Bai}%
  \BibitemOpen
  \href@noop \ \bibinfo {note} {A similar expression for
  the Lyapunov susceptibility has been obtained within the evolution-operator
  formalism in Ref.~\cite{Bai09}.}\BibitemShut {Stop}%
\bibitem [{\citenamefont {Lackovi\'{c}}(1982)}]{convex82}%
  \BibitemOpen
  \bibfield  {author} {\bibinfo {author} {\bibfnamefont {Ivan~B.}\ \bibnamefont
  {Lackovi\'{c}}},\ }\bibfield  {title} {\enquote {\bibinfo {title} {On the
  behaviour of sequences of left and right derivatives of a convergent sequence
  of convex functions},}\ }\href@noop {} {\bibfield  {journal} {\bibinfo
  {journal} {Publikacije Electrotehni\v{c}kog fakulteta. Serija Matematika i
  fizika}\ ,\ \bibinfo {pages} {19}} (\bibinfo {year} {1982})}\BibitemShut
  {NoStop}%
\end{thebibliography}%
\end{document}